\documentclass[a4paper, amsfonts, amssymb, amsmath, reprint, showkeys, nofootinbib, twoside,notitlepage,onecolumn]{revtex4-2}

\def\apj{{ApJ}}                 

\def\mnras{{MNRAS}}             
\def\prd{{Phys.~Rev.~D}}        
\def\prl{{Phys.~Rev.~Lett.}}    
\def\jqsrt{{J.~Quant.~Spec.~Radiat.~Transf.}}

\usepackage{amsmath,amstext}
\usepackage[T1]{fontenc}
\usepackage{amssymb}
\input{epsf}
\usepackage{graphicx}
\usepackage{ae,aecompl}

\usepackage{hyperref}
\usepackage{amsmath}
\usepackage{amssymb}
\usepackage{mathtools}
\usepackage{bm}
\usepackage{cleveref}
\usepackage{tensor}
\usepackage{braket}
\usepackage{enumitem}
\usepackage{mhchem}
\usepackage{cancel}
\usepackage{amsthm}
\usepackage{upgreek}
\usepackage{mathrsfs}
\usepackage{slashed}

\usepackage{color}

\newcommand{\spc}{\quad \quad \quad}

\newcommand{\w}{{\mathfrak{w}}}
\newcommand{\q}{{\mathfrak{q}}}

\def\be{\begin{equation}}
\def\ee{\end{equation}}
\def\beq{\begin{eqnarray}}
\def\eeq{\end{eqnarray}}

\theoremstyle{definition}

\theoremstyle{theorem}

\theoremstyle{corollary}

\begin{document}
\title{Dispersion relations of relativistic radiation hydrodynamics}
\author{ L.~Gavassino}
\affiliation{
Department of Mathematics, Vanderbilt University, Nashville, TN, USA
}

\begin{abstract}
We compute the linearised dispersion relations of shear waves, heat waves, and sound waves in relativistic ``matter+radiation'' fluids with grey absorption opacities. This is done by solving radiation hydrodynamics perturbatively in the ratio ``radiation stress-energy''/``matter stress-energy''. The resulting expressions $\omega \, {=} \, \omega(k)$ accurately describe the hydrodynamic evolution for any $k\, {\in}\, \mathbb{R}$. General features of the dynamics (e.g., covariant stability, propagation speeds, and damping of discontinuities) are argued directly from the analytic form of these dispersion relations.
\end{abstract}

\maketitle

\section{Introduction}\label{Introne}
\vspace{-0.2cm}

Every fluid is filled with a gas of thermal photons. Such photons must participate in the motion of the fluid, since they can exchange energy and momentum with the other constituent particles. On Earth, the backreaction that these photons exert on the motion of matter is usually small, and may be neglected. However, if one considers hotter environments, such as stars (especially the most massive ones\footnote{\label{footnes1}The dimensionless number that quantifies the importance of radiation-hydrodynamic effects in an astrophysical gas or plasma is the ratio $\mathcal{R}=\text{``Radiation pressure''}/\text{``Gas pressure''} \sim T^4/(nT)$, where $T$ is the temperature and $n$ the gas number density (natural units are adopted, so that $c{=}\hbar{=}k_B{=}1$). For atmospheric air, one has $\mathcal{R}\sim 10^{-11}$. In a star with mass $M$ and composition $\mu$, the ratio $\mathcal{R}$ is related to Eddington's $\beta$-parameter by the identity $\beta=(1{+}\mathcal{R})^{-1}$, and we have $\mathcal{R}(1{+}\mathcal{R})^3  \approx 0.003 \, \mu^4 (M/M_\odot)^2$ \cite[\S 5.6]{Prialnik2009}.}), photons can become the main engine driving a flow. The branch of fluid mechanics that studies the impact of thermal photons on the macroscopic motion of fluids is called ``radiation hydrodynamics'', and its applications span the whole field of relativistic astrophysics \cite{Pomraning1973,mihalas_book,CastorRadiationBook_2004,Thomas1930,Weinberg1971,UdeyIsrael1982,Thorne1981,AnileRadiazion1992,Farris2008,Sadowski2013}.

From a fundamental physics perspective, radiation hydrodynamics is also particularly interesting in that it is a hybrid model, where matter is governed by fluid mechanics (relativistic or not), and radiation is governed by relativistic kinetic theory \cite{Groot1980RelativisticKT}. Thus, solving the equations of radiation hydrodynamics requires solving the hydrodynamic equations for the matter fields, coupled with the full Boltzmann equation for the radiation distribution function $f(x^\alpha,p^\alpha)$, which counts how many photons are found at a spacetime location $x^\alpha$ and occupy a state with four-momentum $p^\alpha$ \cite[\S 22.6]{MTW_book}. This gives rise to a very rich phenomenology, which is usually not observed within hydrodynamics or ideal-gas kinetic theory alone, but explicitly involves the combination of the two. 


In this article, we study the collective excitations of radiation-hydrodynamic systems in full special relativity. Specifically, we linearize the equations of motion about global equilibrium, and we compute the dispersion relations of those quasi-normal modes that involve a fluctuation of the matter component. These are the so-called ``hydrodynamic modes'' \cite{McLennan1965,Kurkela:2017xis,Grozdanov2019,Romatschke:2015gic,GavassinoNonHydro2022,GavassinoGapless2024rck,GavassinoChapmanEnskog2024xwf}, and they are 5 in total, which can be classified as follows: 2 shear waves, 1 heat wave, and 2 sound waves. We will show that, if \textbf{(a)} scattering is neglected, \textbf{(b)} the opacity is grey, and \textbf{(c)} the stress-energy tensor of radiation is small compared to that of matter (e.g. $\mathcal{R}\, {\lesssim}\, 0.01$, see footnote \ref{footnes1}), then the linearised dispersion relations are ($\omega\,{=}\,$``frequency''$+i \times$ ``growing rate''$\,{\in}\, \mathbb{C}$, $k\,{=}\,$``wavenumber''$\,{\in}\, \mathbb{R}$, $\tau \, {=}\,$``photon mean free path''$\, >0$):
\begin{flalign}\label{mainformula}
\text{Shear waves:} \quad \quad \omega(k) =-i \dfrac{15D_s}{2\tau^2} \bigg[ \dfrac{2}{3}+\dfrac{1}{(k\tau)^2}-\bigg( 1{+}\dfrac{1}{(k\tau)^2} \bigg) \dfrac{\arctan(k\tau)}{k\tau}  \bigg] \, , &&
\end{flalign}
\begin{flalign}\label{mainformula2}
\text{Heat waves:} \quad \quad \, \, \omega(k) =-i \dfrac{3D_h}{\tau^2} \bigg[1-\dfrac{\arctan(k\tau)}{k\tau} \bigg]\, , &&
\end{flalign}
\begin{flalign}\label{mainformula3}
\text{Sound waves:} \quad \! \quad  \omega(k) = c_s k -i \dfrac{15 D_s}{2\tau^2} \bigg[ \dfrac{1}{3}-\dfrac{1{-}ic_s k\tau}{(k\tau)^2} +\dfrac{(1{-}ic_s k\tau)^2}{(k\tau)^3} \arctan\bigg(\dfrac{k\tau}{1{-}ic_s k\tau} \bigg)   \bigg] \, , &&
\end{flalign}
where $D_s$ and $D_h$ are two effective (radiation-induced) diffusivities, and $c_s$ is the sound speed of matter alone\footnote{\label{kpvanishes}Equation \eqref{mainformula3} is derived under the additional assumption that the isobaric thermal expansivity of the matter component vanishes.}. Then, we will discuss the mathematical properties and physical implications of \eqref{mainformula}-\eqref{mainformula3}.


Some of the results of this article are not entirely new. For example, equation \eqref{mainformula2} for heat waves was computed for the first time by \citet{Spiegel1957}. However, our work differs from the previous literature in two aspects. First, the matter component is evolved self-consistently, and no dynamic constraint is assumed, while previous analyses make restrictive assumptions on the flow. For example, \citet{Spiegel1957} holds the matter component at rest and, as a result, their coefficient $D_h$ is, strictly speaking, incorrect. Secondly, our treatment of both radiation and matter is fully relativistic. This is especially important for sound waves, whose speed may become comparable to the speed of radiation itself. Furthermore, in relativity, accelerations are sources of heat \cite{Eckart40,GavassinoLyapunov_2020}, which increases the damping rate of sound waves.

Throughout the article, we work in Minkowski spacetime, with metric signature $(-,+,+,+)$, and we adopt natural\ units, such that $c=\hbar=k_B=1$. Greek indices run from 0 to 3 (with $x^0=t$), while Latin indices run from 1 to 3.

\section{Derivation of the disperison relations}\label{derivZia}

In this section, we derive the dispersion relations \eqref{mainformula}-\eqref{mainformula3} by solving the equations of radiation hydrodynamics.

As explained in the introduction, we are dealing with a ``radiation+matter'' fluid, i.e. an interacting mixture of two distinct physical components: a material medium $M$ with negligibly short mean free path, plus a radiation gas $R$ of photons with finite mean free path $\tau>0$. 


\subsection{The Matter (M) sector}\label{Mmmmm}

The medium $M$ is an ideal fluid in local thermodynamic equilibrium (because its relaxation times are instantaneous), with a well-defined flow velocity field $u^\mu$, and with thermodynamic fields $\{\rho,P,s,n,T,\mu \}$, representing respectively energy density, pressure, entropy density, baryon density, temperature, and baryon chemical potential. These fields are related by usual thermodynamic identities, like the first law of thermodynamics, the Gibbs-Duhem relation, and the Euler relation, respectively:
\begin{equation}
\begin{split}
d\rho ={}& T \, ds +\mu \, dn \, , \\
dP ={}& s \, dT +n \, d\mu \, , \\
\rho{+}P ={}& Ts +\mu n \, . \\
\end{split}
\end{equation}
By assumption, the stress-energy tensor and baryon current associated with $M$ are those of an ideal fluid:
\begin{equation}
\begin{split}
T_M^{\mu \nu}={}& (\rho{+}P)u^\mu u^\nu +P g^{\mu \nu} \, ,\\
J_M^\mu ={}& n u^\mu \, . \\
\end{split}
\end{equation}

\subsection{The Radiation (R) sector}

The kinetic state of the radiation component $R$ is fully characterized by its invariant distribution function $f(x^\alpha,p^\alpha)$ \cite{MTW_book}, where $p^\alpha$ is the photon four-momentum (with $p^\alpha p_\alpha=0$).
We assume that matter-radiation interactions occur solely through the absorption and emission of photons by the medium $M$ [see section \ref{Introne}, assumption \textbf{(a)}]. Then, recalling that $M$ is in local equilibrium, the radiative Boltzmann equation for photons is given by \cite[\S 92]{mihalas_book},
\begin{equation}\label{Boltzmann}
p^\mu \partial_\mu f = \dfrac{p^\mu u_\mu}{\tau} (f-f_{\text{eq}}) \, ,
\end{equation}
where we recall that $u^\mu$ is the flow velocity of the medium $M$. For simplicity, we make the standard ``grey opacity'' assumption [see section \ref{Introne}, assumption \textbf{(b)}], according to which $\tau$ is a constant, independent of $p^\mu$. Note that the right-hand side of \eqref{Boltzmann} can be divided into two parts. The part ``$p^\mu u_\mu f/\tau $'' is the sink term describing absorption processes, while the part ``$-p^\mu u_\mu f_{\text{eq}}/\tau$'' is the source term describing emission processes. As usual, the second term is related to the first by the Kirchhoff-Planck relation \cite[\S 72]{mihalas_book}, according to which the source and the sink must cancel out when radiation is in thermal equilibrium with matter, namely when $f$ coincides with the Planckian distribution
\begin{equation}\label{equilub}
    f_\text{eq}= \dfrac{1}{e^{-\beta_\nu p^\nu} -1}   \spc (\text{with }\beta_\nu=u_\nu/T) \, ,
\end{equation}
where the field $\beta^\nu$ is the ``inverse-temperature four-vector'' \cite{Israel_Stewart_1979,BecattiniBeta2016,GavassinoTermometri} of the medium. 
The stress-energy tensor and baryon current of radiation can be expressed in terms of the kinetic distribution function $f$ as follows \cite[\S 91]{mihalas_book}\cite[\S 6.3]{CastorRadiationBook_2004}:
\begin{equation}\label{stressOne}
\begin{split}
T_R^{\mu \nu} ={}& 2 \int \dfrac{d^3 p}{(2\pi)^3 p^0} \, p^\mu p^\nu \, f \, , \\
J_R^\mu ={}& 0 \, .
\end{split}
\end{equation}
The factor $2$ in the definition of $T_R^{\mu \nu}$ accounts for the spin degeneracy\footnote{We define $f$ as the average occupation number of single-photon states, and we assume that both spin polarizations have equal occupation.}. The baryon current $J_R^\mu$ of radiation vanishes because the photon is its own antiparticle, and thus cannot carry conserved quantum numbers. 

\subsection{Linearised equations of motion}

The dynamical degrees of freedom of radiation hydrodynamics are $\Psi=\{\rho,u^\mu,n,f \}$. Therefore, the conservation laws $\partial_\mu (T_M^{\mu \nu}{+}T_R^{\mu \nu})=0$ and $\partial_\mu (J_M^\mu {+}J_R^\mu)=0$, plus the Boltzmann equation \eqref{Boltzmann}, are enough to fully determine the evolution. We linearized all these equations of motion around a uniform equilibrium state with $u^\mu=(1,0,0,0)$, $\rho =\text{const}$, $n=\text{const}$, and $f=f_\text{eq}$. The result is reported below:
\begin{flalign}\label{settatntuno}
\boxed{\rho^{\textcolor{white}{1}}} \quad \partial_t \delta \rho+(\rho{+}P)\partial_j \delta u^j + \int \dfrac{2d^3 p}{(2\pi)^3} p^\mu \partial_\mu \delta f =0 \, , &&
\end{flalign}
\begin{flalign}\label{settantadue}
\boxed{u^1} \quad (\rho{+}P)\partial_t \delta u^1+\partial_1 \delta P + \int \dfrac{2d^3 p}{(2\pi)^3} \dfrac{p^1}{p^0} p^\mu \partial_\mu \delta f =0 \, , &&
\end{flalign}
\begin{flalign}\label{settantatre}
\boxed{u^2} \quad (\rho{+}P)\partial_t \delta u^2+\partial_2 \delta P + \int \dfrac{2d^3 p}{(2\pi)^3} \dfrac{p^2}{p^0} p^\mu \partial_\mu \delta f =0 \, , &&
\end{flalign}
\begin{flalign}\label{settantquattro}
\boxed{u^3} \quad (\rho{+}P)\partial_t \delta u^3+\partial_3 \delta P + \int \dfrac{2d^3 p}{(2\pi)^3} \dfrac{p^3}{p^0} p^\mu \partial_\mu \delta f =0 \, , &&
\end{flalign}
\begin{flalign}\label{settantcinque}
\boxed{n^{\textcolor{white}{1}}} \quad  \partial_t \delta n +n\partial_j \delta u^j =0 \, , &&
\end{flalign}
\begin{flalign}\label{settantasei}
\boxed{f^{\textcolor{white}{1}}} \quad \dfrac{\tau}{p^0} p^\mu \partial_\mu \delta f + \delta f =f_\text{eq}(1{+}f_{\text{eq}}) \, p^\nu \delta \beta_\nu \, , &&
\end{flalign}
where ``$\delta \Psi$'' is the linear perturbation to $\Psi$, and the boxes before the equations serve to keep track of which degree of freedom is evolved by which equation. We search for solutions in the form of sinusoidal waves that propagate in direction 1. This just means that we assume that all quantities have a spacetime dependence of the form $e^{i(kx^1-\omega t)}$, with $k \in \mathbb{R}$ and $\omega \in \mathbb{C}$. With this assumption, equation \eqref{settantasei} reduces to
\begin{equation}\label{df}
\delta f= \dfrac{f_\text{eq}(1+f_{\text{eq}}) p^\nu \delta \beta_\nu}{1-i\omega \tau +ik\tau p^1/p^0} \, .
\end{equation}
Adopting the decomposition $p^\mu =p^0 \Omega^\mu$, with $\Omega^0=\Omega^j \Omega_j=1$, equations \eqref{settatntuno}-\eqref{settantcinque} become
\begin{flalign}\label{2settatntuno}
\boxed{\rho^{\textcolor{white}{1}}} \quad  \omega \delta \rho-k (\rho{+}P) \delta u^1 + \dfrac{\pi T^5}{15} \int_{\mathcal{S}^2} \dfrac{(\omega{-}k \Omega^1) \Omega^\nu \delta \beta_\nu}{1{-}i \omega \tau {+} ik\tau \Omega^1} \, d^2 \Omega =0 \, , &&
\end{flalign}
\begin{flalign}\label{2settantadue}
\boxed{u^1} \quad  \omega (\rho{+}P) \delta u^1-k \delta P + \dfrac{\pi T^5}{15} \int_{\mathcal{S}^2} \dfrac{(\omega{-}k \Omega^1) \Omega^1 \Omega^\nu \delta \beta_\nu}{1{-}i \omega \tau {+} ik\tau \Omega^1} \, d^2 \Omega  =0 \, , &&
\end{flalign}
\begin{flalign}\label{2settantatre}
\boxed{u^2} \quad  \omega (\rho{+}P) \delta u^2 + \dfrac{\pi T^5}{15} \int_{\mathcal{S}^2} \dfrac{(\omega{-}k \Omega^1)\Omega^2 \Omega^\nu \delta \beta_\nu}{1{-}i \omega \tau {+} ik\tau \Omega^1} \, d^2 \Omega  =0 \, , &&
\end{flalign}
\begin{flalign}\label{2settantquattro}
\boxed{u^3} \quad \omega(\rho{+}P) \delta u^3 + \dfrac{\pi T^5}{15} \int_{\mathcal{S}^2} \dfrac{(\omega{-}k \Omega^1) \Omega^3 \Omega^\nu \delta \beta_\nu}{1{-}i \omega \tau {+} ik\tau \Omega^1} \, d^2 \Omega  =0 \, , &&
\end{flalign}
\begin{flalign}\label{2settantcinque}
\boxed{n^{\textcolor{white}{1}}} \quad   \dfrac{\delta n}{n} = \dfrac{k}{\omega}  \delta u^1  \, , &&
\end{flalign}
where we have evaluated the integral in the variable $p^0$. The above equations are exact (within the model assumptions), and they can be used to compute all the (gapless \cite{GavassinoNonHydro2022}) dispersion relations of radiation hydrodynamics. The integration element $d^2 \Omega$ is the solid angle in spherical coordinates.

\subsection{Shear waves}\label{shearone}

Let us solve equations \eqref{2settatntuno}-\eqref{2settantcinque} for a transversal wave that fluctuates in direction 3, namely for a configuration such that $\delta u^1=\delta u^2=0 \neq \delta u^3$. Then, equation \eqref{2settantcinque} immediately gives $\delta n=0$. Furthermore, $\delta \beta_\nu =T^{-1}(\delta T/T,0,0,\delta u^3)$. Therefore, we have 
\begin{flalign}\label{3settatntuno}
\boxed{\rho^{\textcolor{white}{1}}} \quad  \omega \delta \rho + \dfrac{\pi T^4}{15} \int_{\mathcal{S}^2} \dfrac{(\omega{-}k \Omega^1) (\delta T/T{+}\Omega^3 \delta u_3)}{1{-}i \omega \tau {+} ik\tau \Omega^1} \, d^2 \Omega =0 \, , &&
\end{flalign}
\begin{flalign}\label{3settantadue}
\boxed{u^1} \quad  -k \delta P + \dfrac{\pi T^4}{15} \int_{\mathcal{S}^2} \dfrac{(\omega{-}k \Omega^1) \Omega^1 (\delta T/T{+}\Omega^3 \delta u_3)}{1{-}i \omega \tau {+} ik\tau \Omega^1} \, d^2 \Omega  =0 \, , &&
\end{flalign}
\begin{flalign}\label{3settantatre}
\boxed{u^2} \quad   \dfrac{\pi T^4}{15} \int_{\mathcal{S}^2} \dfrac{(\omega{-}k \Omega^1)\Omega^2 (\delta T/T{+}\Omega^3 \delta u_3)}{1{-}i \omega \tau {+} ik\tau \Omega^1} \, d^2 \Omega  =0 \, , &&
\end{flalign}
\begin{flalign}\label{3settantquattro}
\boxed{u^3} \quad \omega(\rho{+}P) \delta u^3 + \dfrac{\pi T^4}{15} \int_{\mathcal{S}^2} \dfrac{(\omega{-}k \Omega^1) \Omega^3 (\delta T/T{+}\Omega^3 \delta u_3)}{1{-}i \omega \tau {+} ik\tau \Omega^1} \, d^2 \Omega  =0 \, . &&
\end{flalign}
Equation \eqref{3settantatre} is an identity ``$\, 0{=}0\,$'' due to the factor $\Omega^2$ (the second component of $\Omega^j$) in the integrand, which averages to zero when integrated over the sphere. For a similar reason, the term $\Omega^3 \delta u_3$ averages to zero in equations \eqref{3settatntuno} and \eqref{3settantadue}, which are simultaneously satisfied only when $\delta T=0$ (which implies $\delta \rho{=}\delta P{=}0$, since $\delta n$ also vanishes). Hence, we are left only with equation \eqref{3settantquattro}, where $\delta u^3$ cancels out. Introducing the real dimensionless quantities $\Gamma=-i\omega \tau$ and $\q=k\tau$, we obtain the following \textit{exact} dispersion relation, expressed in an implicit form:
\begin{equation}\label{www}
\Gamma +\dfrac{2\pi^2 T^4}{15(\rho{+}P)} \bigg[ \dfrac{2}{3}+\dfrac{1{+}\Gamma}{\q^2}-\bigg( 1{+}\dfrac{(1{+}\Gamma)^2}{\q^2} \bigg) \dfrac{1}{\q} \arctan\bigg( \dfrac{\q}{1{+}\Gamma}\bigg)  \bigg]=0  \, .
\end{equation}
This relation can be converted into an exact parametric expression $\{k(r),\omega(r)\}$, see \cite{GavassinoShearRadiation2024jej}. Here, we will examine the limit of \eqref{www} when $T_R^{00}/T_M^{00}\ll 1$ [see section \ref{Introne}, assumption \textbf{(c)}], which will give us a simple formula for $\omega(k)$. To this end, we first fix the value of $\q {\in} \mathbb{R}\backslash \{0\}$, and treat it as a constant. Then, we define a small free parameter
\begin{equation}\label{lambo}
    \lambda = \dfrac{2\pi^2 T^4}{15(\rho{+}P)} \sim \dfrac{T_R^{00}}{T_M^{00}} \, .
\end{equation}
This allows us to interpret $\Gamma$ a function of $\lambda$, which may be Taylor expanded in $\lambda$, namely $\Gamma(\lambda)=\Gamma(0)+\lambda  \Gamma'(0)+\mathcal{O}(\lambda^2)$. To compute $\Gamma(0)$ and $\Gamma'(0)$, we only need to regard \eqref{www} as an implicit function $F(\lambda,\Gamma(\lambda))=0$. Setting $\lambda=0$, we immediately obtain $\Gamma(0)=0$. Differentiating in $\lambda$ at $0$, we obtain 
\begin{equation}
\Gamma'(0)=-\dfrac{\partial_\lambda F(0,0)}{\partial_\Gamma F(0,0)}= -\bigg[ \dfrac{2}{3}+\dfrac{1}{\q^2}-\bigg( 1{+}\dfrac{1}{\q^2} \bigg) \dfrac{\arctan(\q)}{\q}  \bigg] \, . 
\end{equation}
Thus, we have an approximate formula for the frequency: $\omega=i\lambda \Gamma'(0)/\tau+\mathcal{O}(\lambda^2)$. Explicitly, this reads
\begin{equation}\label{seriaccio}
\begin{split}
\omega ={}& -i  \dfrac{2\pi^2 T^4}{15(\rho{+}P)\tau} \bigg[ \dfrac{2}{3}+\dfrac{1}{(k\tau)^2}-\bigg( 1{+}\dfrac{1}{(k\tau)^2} \bigg) \dfrac{\arctan(k\tau)}{k\tau}  \bigg] +\mathcal{O}(\lambda^2) \\
={}& -i  \dfrac{4\pi^2 T^4}{225(\rho{+}P)\tau} \bigg[(k\tau)^2 -\dfrac{3(k\tau)^4}{7} + \dfrac{5(k\tau)^6}{21} -...\bigg] +\mathcal{O}(\lambda^2) \, , \\
\end{split}
\end{equation}
where the series in the second line converges for $|k\tau|<1$. Direct inspection of the first term of such series allows us to read out the diffusion coefficient \cite{HellerBounds2023} of shear waves, namely
\begin{equation}\label{diffuzio}
D_s = \dfrac{4\pi^2 T^4 \tau}{225(\rho{+}P)} \, ,
\end{equation}
and this finally leads us to the formula we were looking for (note that $ D_s/\tau \sim \lambda$):
\begin{equation}
\omega =-i \dfrac{15D_s}{2\tau^2} \bigg[ \dfrac{2}{3}+\dfrac{1}{(k\tau)^2}-\bigg( 1{+}\dfrac{1}{(k\tau)^2} \bigg) \dfrac{\arctan(k\tau)}{k\tau}  \bigg] +\mathcal{O}(D_s^2/\tau^2) \, .
\end{equation}
We stress that this formula is a good approximation of kinetic theory for arbitrary values of $k \in \mathbb{R}$.

We also remark that \eqref{mainformula} coincides with the corresponding formula of \cite{GavassinoShearRadiation2024jej}. This is reassuring, since the analysis of \cite{GavassinoShearRadiation2024jej} was based on a purely geometrical argument, while here we have solved all the equations explicitly.


\subsection{Heat waves}\label{heattoneee}

Let us now derive the dispersion relation of heat waves. To this end, we go back to the original system \eqref{2settatntuno}-\eqref{2settantcinque}. However, this time, we consider a longitudinal wave, i.e. a sinusoidal perturbation with vanishing tranversal velocities: $\delta u^2=\delta u^3=0$. Then, the fluctuation to the inverse-temperature four-vector is just $\delta \beta_\nu =T^{-1}(\delta T/T,\delta u^1,0,0)$, and equations \eqref{2settantatre}-\eqref{2settantquattro}  become trivial identities ``$\, 0{=}0\,$'' (again, the integrals vanish because the components $\Omega^2$ and $\Omega^3$ average to zero when integrated over all angles). With the aid of the first law of thermodynamics \cite{Hishcock1983,MTW_book}
\begin{equation}\label{griguzzone}
\begin{split}
\delta \rho ={}& \dfrac{\rho{+}P}{n} \,  \delta n + nT \, \delta \mathfrak{s} \, , \\
\end{split}
\end{equation}
where $\mathfrak{s}$ is the specific entropy of the fluid component, we can rewrite the system \eqref{2settatntuno}-\eqref{2settantcinque} as follows:
\begin{flalign}\label{7settatntuno}
\boxed{\rho^{\textcolor{white}{1}}} \quad  \omega \, \delta \mathfrak{s} + \dfrac{\pi T^3}{15 \, n} \int_{\mathcal{S}^2} \dfrac{(\omega{-}k \Omega^1) (\delta T/T{+}\Omega^1 \delta u_1)}{1{-}i \omega \tau {+} ik\tau \Omega^1} \, d^2 \Omega =0 \, , &&
\end{flalign}
\begin{flalign}\label{7settantadue}
\boxed{u^1} \quad  \omega (\rho{+}P) \delta u^1-k \delta P + \dfrac{\pi T^4}{15} \int_{\mathcal{S}^2} \dfrac{(\omega{-}k \Omega^1) \Omega^1 (\delta T/T{+}\Omega^1 \delta u_1)}{1{-}i \omega \tau {+} ik\tau \Omega^1} \, d^2 \Omega  =0 \, , &&
\end{flalign}
\begin{flalign}\label{7settantcinque}
\boxed{n^{\textcolor{white}{1}}} \quad     \dfrac{\delta n}{n} = \dfrac{k}{\omega}  \delta u^1  \, . &&
\end{flalign}
This can be viewed as a system in the fluid variables $\{\delta \mathfrak{s},\delta P,\delta u^1 \}$, if one recalls the thermodynamic identities \cite{GavassinoNonHydro2022}
\begin{equation}\label{gringuzzino}
\dfrac{\delta T}{T} =  \dfrac{\kappa_p}{nc_p} \, \delta P + \dfrac{\delta \mathfrak{s}}{c_p} \, , \spc 
\dfrac{\delta n}{n} = \dfrac{\delta P}{c_s^2 (\rho{+}P)} - \dfrac{T\kappa_p}{c_p} \, \delta \mathfrak{s} \, , 
\end{equation}
where $c_p$, $c_s^2$ and $\kappa_p$ are respectively the specific heat at constant pressure, the adiabatic speed of sound squared, and the isobaric thermal expansivity (a.k.a. expansion coefficient) of the matter component. Thus, introducing again the dimensionless quantities $\Gamma=-i\omega \tau$ and $\q=k\tau$, and defining a new small parameter [see section \ref{Introne}, assumption \textbf{(c)}]
\begin{equation}\label{nuino}
    \nu = \dfrac{4\pi^2 T^3}{15 \, nc_p} \sim \dfrac{T_R^{11}}{T_M^{11}}  \, ,
\end{equation}
we can rewrite the system \eqref{7settatntuno}-\eqref{7settantcinque} in the following (exact) form:
\begin{flalign}\label{8settatntuno}
\boxed{\rho^{\textcolor{white}{1}}} \quad  \Gamma  =- \nu \int_{-1}^1 \dfrac{\Gamma{+}i\q \xi}{1{+}\Gamma {+} i \q \xi} \bigg[1+ \dfrac{\kappa_p}{n}  \delta P  +\xi c_p \delta u^1   \bigg]  \, \dfrac{d\xi}{2}  \, , &&
\end{flalign}
\begin{flalign}\label{8settantadue}
\boxed{u^1} \quad   \bigg[1{+} \dfrac{\Gamma^2}{c_s^2 \q^2}\bigg]\delta P - \dfrac{\Gamma^2}{\q^2} \dfrac{T\kappa_p }{c_p} (\rho{+}P) =i \dfrac{nT \nu}{\q} \int_{-1}^1 \dfrac{(\Gamma{+} i\q \xi) \xi}{1{+}\Gamma {+} i \q \xi} \bigg[ 1+\dfrac{\kappa_p}{n}  \delta P  +\xi c_p \delta u^1   \bigg] \, \dfrac{d\xi}{2}   \, , &&
\end{flalign}
\begin{flalign}\label{8settantcinque}
\boxed{n^{\textcolor{white}{1}}} \quad     \delta u^1 = i\dfrac{\Gamma}{\q} \bigg[\dfrac{\delta P}{c_s^2 (\rho{+}P)} - \dfrac{T\kappa_p}{c_p} \bigg]    \, , &&
\end{flalign}
where we have employed the linearity of the equations to formally set $\delta \mathfrak{s}=1$.\footnote{\label{uonno}In this way, we are also forcing the system to give us the heat wave as our only solution. In fact, the system \eqref{7settatntuno}-\eqref{7settantcinque} possesses three linearly independent solutions, where two solutions are sound waves, and the remaining one is the heat wave. In the limit as $\nu \rightarrow 0$, the sound waves become adiabatic, and thus have $\delta \mathfrak{s}=0$. Therefore, if we set $\delta \mathfrak{s}=1$, and Taylor-expand the system for small $\nu$, we automatically rule out the sound waves.} Similarly to what we did in the previous subsection, we fix the value of $\q \in \mathbb{R}\backslash \{0\}$ (which may be large), and consider the list of functions $X(\nu)=\{\Gamma(\nu),\delta P(\nu),\delta u^1(\nu)\}$. We expand all such functions to first order in $\nu$, i.e. $X(\nu)=X(0)+\nu X'(0)+\mathcal{O}(\nu^2)$. The zeroth order is straightforward: equation \eqref{8settatntuno} gives $\Gamma(0)=0$, equation \eqref{8settantadue} gives $\delta P(0)=0$, and equation \eqref{8settantcinque} gives $\delta u^1(0)=0$, which is what we expect from a heat wave in an ideal fluid. At first order, we find
\newpage
\begin{flalign}\label{9settatntuno}
\boxed{\rho^{\textcolor{white}{1}}} \quad  \Gamma'(0)  =-i\q  \int_{-1}^1 \dfrac{ \xi}{1 {+} i \q \xi}  \, \dfrac{d\xi}{2} =- \bigg( 1- \dfrac{\arctan \q }{\q}\bigg)  \, , &&
\end{flalign}
\begin{flalign}\label{9settantadue}
\boxed{u^1} \quad    \delta P'(0) = -nT \int_{-1}^1 \dfrac{ \xi^2}{1 {+} i \q \xi}  \, \dfrac{d\xi}{2} = -\dfrac{nT}{\q^2} \bigg( 1- \dfrac{\arctan \q}{\q} \bigg)  \, , &&
\end{flalign}
\begin{flalign}\label{9settantcinque}
\boxed{n^{\textcolor{white}{1}}} \quad     \delta {u^1}'(0) = -i \dfrac{T\kappa_p}{\q c_p}\Gamma'(0) =  i \dfrac{T\kappa_p}{\q c_p} \bigg( 1- \dfrac{\arctan \q }{\q}\bigg)   \, . &&
\end{flalign}
Thus, we finally obtain
\begin{equation}\label{capolavoro}
\begin{split}
\omega ={}& -i  \dfrac{4\pi^2 T^3}{15 \, nc_p \tau} \bigg[1{-} \dfrac{\arctan(k\tau)}{k\tau} \bigg]+\mathcal{O}(\nu^2)= -i  \dfrac{4\pi^2 T^3}{15 \, nc_p \tau} \bigg[\dfrac{(k\tau)^2}{3}- \dfrac{(k\tau)^4}{5}+... \bigg]+\mathcal{O}(\nu^2) \, , \\
\delta P ={}& - \dfrac{4\pi^2 T^4}{15 c_p (k\tau)^2} \bigg[1{-} \dfrac{\arctan(k\tau)}{k\tau} \bigg]+\mathcal{O}(\nu^2) =- \dfrac{4\pi^2 T^4}{15 c_p } \bigg[ \dfrac{1}{3}-\dfrac{(k\tau)^2}{5}+... \bigg]+\mathcal{O}(\nu^2) \, , \\
\delta u^1 ={}& i  \dfrac{4\pi^2 T^4 \kappa_p}{15 \, nc_p^2 k \tau} \bigg[1{-} \dfrac{\arctan(k\tau)}{k\tau} \bigg]+\mathcal{O}(\nu^2)= i \dfrac{4\pi^2 T^4 \kappa_p}{15 \, nc_p^2 } \bigg[\dfrac{k\tau}{3}- \dfrac{(k\tau)^3}{5}+... \bigg]+\mathcal{O}(\nu^2) \, . \\
\end{split}
\end{equation}
From the truncation of $\omega(k)$ at order $k^2$, we obtain the formula for the diffusion coefficient of heat waves, which reads
\begin{equation}\label{fourtieight}
D_h= \dfrac{4\pi^2 T^3 \tau}{45 n c_p} \, .
\end{equation}
Similar to the shear case, we see that $D_h/\tau \sim \nu$. Thus, the small-$\nu$ expansion is equivalent to the expansion for small $D_h/\tau$, and we finally obtain
\begin{equation}
\omega= -i \dfrac{3D_h}{\tau^2} \bigg[1-\dfrac{\arctan(k\tau)}{k\tau} \bigg] + \mathcal{O}\big(D_h^2/\tau^2\big) \, ,
\end{equation}
which is what we wanted to prove. Note that, in the formula for the diffusivity coefficient $D_h$, the specific heat $c_p$ is at constant pressure, and \textit{not} at constant volume\footnote{As mentioned in the introduction, equation \eqref{mainformula2} is formally identical to the quasi-static radiation transport equation given in \cite{Spiegel1957}, which is commonly reported in textbooks \cite[\S 100]{mihalas_book}. However, our coefficient $D_h$ differs from that of \cite{Spiegel1957} by the presence of $c_p$ (instead of $c_v$) in the denominator. This difference arises from the fact that we are evolving the velocity perturbation $\delta u^1$ self-consistently, while in the standard literature one just sets $\delta u^1=0$ to simplify the problem.}, which would otherwise be denoted by $c_v$. This distinction is important because we are dealing with fluids (where indeed heat propagates with $c_p$ \cite[\S 50]{landau6}), while it would be irrelevant in solids \cite[\S 32]{landau7}. We also remark that the dispersion relation given above well approximates kinetic theory for all values of $k\in \mathbb{R}$, including the optically thin limit $|k\tau|\gg 1$.

For completeness, let us comment on the physical interpretation of the perturbations to $P$ and $u^1$ provided in \eqref{capolavoro}. The value of $\delta P$ can be calculated using equation \eqref{8settantadue} alone, which is a rearrangement of \eqref{2settantadue}. The latter is just the conservation of linear momentum in the longitudinal direction, $\partial_t \delta T^{01}+\partial_1 \delta T^{11}=0$. Since the ``acceleration'' term $\partial_t \delta T^{01}\propto \omega \delta u^1$ vanishes to first order in $\nu$, equation \eqref{2settantadue} implies that the perturbations to fluid pressure $\delta P$ and to radiation pressure $\delta P_R$ balance each other, i.e. $\delta T^{11}=\delta P+\delta P_R=0$, so that the composite matter-radiation system is kept at rest, in agreement with the discussion of \citet[\S 50]{landau6}. Indeed, for small $\q$, the radiation gas is in local equilibrium with the fluid (i.e. the black-body formulas apply), and we have, to leading order in $\nu$,
\begin{equation}
\delta P = -\delta P_R = -\delta \bigg(\dfrac{\pi^2 T^4}{45} \bigg) = -\dfrac{4\pi^2T^4}{45} \dfrac{\delta T}{T} = -\dfrac{4\pi^2T^4}{45 c_p} \, ,
\end{equation}
which agrees with the second line of \eqref{capolavoro}, in the limit $\q \rightarrow 0$.

The value of $\delta u^1$ was calculated from equation \eqref{8settantcinque}, which is a rearrangement of the continuity equation \eqref{settantcinque}. The reason for this small correction to the flow velocity is simple: While the fluid elements are kept at constant pressure (to first order in $\nu$), their specific entropy $\mathfrak{s}$ changes over time due to heat diffusion. Thus, the baryon density $n$ also varies in time, forcing the fluid elements to expand and contract by an amount that is proportional to the following thermodynamic coefficient:
\begin{equation}
\dfrac{1}{n}\dfrac{\partial n}{\partial \mathfrak{s}}\bigg|_P = \dfrac{1}{n} \dfrac{\partial n}{\partial T}\bigg|_P \dfrac{\partial T}{\partial \mathfrak{s}}\bigg|_P = -\dfrac{T \kappa_p}{c_p} \, ,
\end{equation}
which indeed appears in the formula for $\delta u^1$, see equation \eqref{capolavoro}.

\subsection{Sound waves}\label{sounduzzonzone}

We are only left with the problem of computing the dispersion relations of sound waves. Since these waves are longitudinal, the relevant system of equations is again \eqref{7settatntuno}-\eqref{7settantcinque}. To simplify the calculations, we assume that the expansion coefficient $\kappa_p$ vanishes (see footnote \ref{kpvanishes}). Thus, $\delta T/T=\delta \mathfrak{s}/c_p$ and $\delta n/n=\delta P/[c_s^2(\rho{+}P)]$. Furthermore, we use the linearity of the equations to set $\delta P=1$, which allows us to rule out heat waves (see footnote \ref{uonno}). Then, introducing again the small parameter $\lambda$ defined in \eqref{lambo}, and adopting the notation $\w=\omega \tau$ and $\q=k\tau$, we obtain 
\begin{flalign}\label{j7settatntuno}
\boxed{\rho^{\textcolor{white}{1}}} \quad   \delta \mathfrak{s} =- \dfrac{\lambda}{nT \w}\int_{-1}^1 \dfrac{(\w{-}\q  \xi)}{1{-}i \w {+} i \q \xi}  \bigg[ (\rho{+}P)\dfrac{\delta \mathfrak{s}}{c_p} + \dfrac{\xi \w}{c_s^2 \q} \bigg] \, d \xi  \, , &&
\end{flalign}
\begin{flalign}\label{j7settantadue}
\boxed{u^1} \quad    \w^2=c_s^2 \q^2 - \lambda c_s^2 \q \int_{-1}^1 \dfrac{(\w{-}\q \xi) \xi }{1{-}i \w {+} i \q \xi} \bigg[(\rho{+}P)\dfrac{\delta \mathfrak{s}}{c_p} + \dfrac{\xi \w}{c_s^2 \q} \bigg] \, d\xi  \, , &&
\end{flalign}
As we did in the previous subsections, we fix $\q \in \mathbb{R}\backslash \{0\}$, and view $\delta \mathfrak{s}$ and $\w$ as functions of $\lambda$. At $\lambda=0$, we have $\delta \mathfrak{s}(0)=0$ and $\w(0)=\pm c_s \q$. We consider the ``$+$'' case for clarity. Then, we can take the total derivative of \eqref{j7settatntuno} and \eqref{j7settantadue} with respect to $\lambda$. This allows us to compute $\delta \mathfrak{s}'(0)$ and $\w'(0)$. Below, we report only the formula for the latter:
\begin{flalign}\label{kj7settantadue}
\boxed{u^1} \quad   \w'(0)=\q   \int_{-1}^1 \dfrac{(\xi{-}c_s) \xi^2 }{1{+} i \q (\xi{-}c_s)} \, \dfrac{d\xi}{2} =-i \bigg[ \dfrac{1}{3} -\dfrac{1{-}ic_s \q}{\q^2} + \dfrac{(1{-}ic_s\q)^2}{\q^3} \arctan \bigg( \dfrac{\q}{1{-}ic_s \q} \bigg)  \bigg]  \, . &&
\end{flalign}
Thus, if we write explicitly the Taylor expansion $\w(\lambda)=\w(0)+\lambda \w'(0)+\mathcal{O}(\lambda^2)$, we finally obtain the desired equation,
\begin{equation}\label{grube}
\omega=c_s k -i \dfrac{15 D_s}{2\tau^2} \bigg[ \dfrac{1}{3} -\dfrac{1{-}ic_s k\tau}{(k\tau)^2} + \dfrac{(1{-}ic_sk\tau)^2}{(k\tau)^3} \arctan \bigg( \dfrac{k\tau}{1{-}ic_s k\tau} \bigg)  \bigg]+ \mathcal{O}(D_s^2/\tau^2) \, ,
\end{equation}
where $D_s$ is the diffusion coefficient of shear waves, defined in equation \eqref{diffuzio}. Just like the previous dispersion relations, also the formula above remains a valid approximation of photon kinetic theory at arbitrarily large $k\tau$. However, differently from the previous cases, we have made here the additional assumption that $\kappa_p=0$, which means that the matter component is assumed not to expand when its temperature is raised at constant pressure.

We remark that $c_s$ and $D_s$ do \textit{not} coincide with the speed of sound and damping coefficient of the sound waves. In fact, if we truncate the dispersion relation \eqref{grube} to second order in $k\tau$, we indeed obtain the usual sound-type long-wavelength expansion $\omega=c_s^{\text{tot}}k-iD_a k^2 +\mathcal{O}(k^2)$, but $c_s^\text{tot} \neq c_s$ and $D_a \neq D_s$. The zeroth-order transport coefficient $c_s^{\text{tot}}$ is actually the ``conglomerate'' speed of sound (i.e. the speed of sound of the composite ``matter+radiation'' fluid), while the first-order transport coefficient $D_a$ is the diffusivity of acoustic waves. The explicit formulas are, respectively,
\begin{equation}\label{cstotdaa}
\begin{split}
c_s^{\text{tot}}={}& c_s \bigg( 1-\dfrac{5D_s}{2\tau}\bigg) \, , \\
D_a ={}& \dfrac{D_s}{2}(3+5c_s^2) \, .\\
\end{split}
\end{equation}
Let us verify explicitly that $c_s^\text{tot}$ is indeed the conglomerate speed of sound that we would obtain from thermodynamics alone, treating photons as ``honorary material particles'' \cite{mihalas_book}. To this end, we need to consider a composite ``matter+radiation'' system in thermal equilibrium, whose energy, pressure, and entropy are the sums of the matter and radiation parts, e.g. $\rho_\text{tot}=\rho+T_R^{00}\big(f_\text{eq}(T)\big)$. Then, with the aid of \eqref{griguzzone}, \eqref{gringuzzino}, and \eqref{nuino}, and defined the imperfect differentials $\slashed{d}\mathcal{Q}=Tnd\mathfrak{s}$ and $\slashed{d}\mathcal{W}=(\rho{+}P)dn/n$, we find that
\begin{equation}\label{grienz}
\begin{split}
d\rho_\text{tot}={} &  (1{+}\nu) \,  \slashed{d}\mathcal{Q}  + \slashed{d}\mathcal{W} \, , \\
dP_\text{tot}={} & \nu  \,  \slashed{d}\mathcal{Q}/3 + c_s^2 \slashed{d}\mathcal{W} \, , \\
Tn d\mathfrak{s}_\text{tot}={} & (1{+}\nu)\slashed{d}\mathcal{Q} - 5D_s  \slashed{d}\mathcal{W}/\tau  \, , \\
\end{split}
\end{equation}
where we recall that $\kappa_p{=}0$ by assumption. Hence, the speed of sound of the composite fluid is (recall that $D_s/\tau {\sim} \lambda$)
\begin{equation}
c_s^\text{tot} =\bigg( \dfrac{\partial P_\text{tot}}{\partial \rho_\text{tot}}\bigg|_{\mathfrak{s}_\text{tot}}\bigg)^{1/2}=\bigg( \dfrac{c_s^2 +\frac{5 \nu D_s}{3\tau (1{+}\nu)}}{1+\frac{5D_s}{\tau}}\bigg)^{1/2} \stackrel{\nu,\lambda \rightarrow 0}{=}  c_s \bigg(1-\dfrac{5D_s}{2\tau} \bigg) +\mathcal{O}[(\nu {+}\lambda)^2]\, ,
\end{equation}
which is what we wanted to prove.

\section{Optically thick and optically thin limits}

Now that the dispersion relations \eqref{mainformula}, \eqref{mainformula2} and \eqref{mainformula3} have been formally derived, let us discuss their limiting behavior as $k\tau \rightarrow 0$ (``optically thick'' limit) and $k\tau \rightarrow \infty$ (``optically thin'' limit)\footnote{Here, the ``optical thickness'' refers to the geometry of the perturbation $\delta \Psi$, and not that of the background state $\Psi$. In fact, the latter is an infinite uniform fluid, so its optical thickness is infinitely large.}.

\subsection{Optically thick limit of diffusive modes}

We have already shown through equations \eqref{seriaccio} and \eqref{capolavoro} that, for small $k\tau$, the dispersion relations of shear and heat waves acquire the standard diffusive form $\omega = -iD k^2+\mathcal{O}(k^4\tau^4)$. Let us now confirm that the effective shear viscosity $\eta$ and the effective heat conductivity $\kappa$ that one obtains in this limit agree with those provided by \citet{Weinberg1971}, who treated the whole ``matter+radiation'' system as an effective viscous fluid. 

Let us first compute the shear viscosity coefficient $\eta$.
To this end, we recall that the evolution equation of shear waves in a relativistic viscous fluid (governed by relativistic Navier-Stokes \cite{Weinberg1971,Eckart40}) is
\begin{equation}
\partial_t \delta T^{03} +\partial_1 \delta T^{13}= (\rho{+}P)\partial_t \delta u_3 -\eta \partial_1^2 \delta u_3=0 \, .
\end{equation}
We note that this has indeed the form of a diffusion equation, with shear diffusivity coefficient $D_s{=}\eta/(\rho{+}P)$ \cite[\S IId]{Weinberg1971}. Thus, if we multiply both sides of \eqref{diffuzio} by $\rho{+}P$, we obtain an effective (Navier-Stokes-type) shear viscosity coefficient,
\begin{equation}\label{etuzzo}
\eta = \dfrac{4\pi^2}{225} T^4 \tau = \dfrac{4}{15} a T^4 \tau  \, ,
\end{equation}
where $a=\pi^2/15$ is the usual radiation constant \cite{rezzolla_book}. Equation \eqref{etuzzo} agrees with \cite{Weinberg1971,Misner1968,Rebetzky1990}.

Let us now compute the heat conductivity coefficient $\kappa$. This time, it is enough to recall that the heat diffusivity coefficient (as provided in textbooks \cite[\S 50]{landau6}) is $D_h =\kappa/(nc_p)$. Thus, multiplying both side of \eqref{fourtieight} by $nc_p$, we obtain the well-known formula for the radiative heat conductivity \cite{NovikovThorne1973}, in agreement with \cite{Weinberg1971}:
\begin{equation}\label{kappuzzo}
\kappa = \dfrac{4\pi^2}{45} T^3 \tau = \dfrac{4}{3} a T^3 \tau  \, .
\end{equation}

\subsection{Optically thick limit of sound modes}

In section \ref{sounduzzonzone}, we showed that, for small values of $k\tau$, the dispersion relation of sound waves acquires the usual form $\omega=c_s^\text{tot} k-iD_a k^2+\mathcal{O}(k^3\tau^3)$, where $c_s^\text{tot}$ is the speed of sound of the total ``matter+radiation'' in local thermodynamic equilibrium. Thus, to confirm that \eqref{mainformula3} has the expected optically thick limit, we only need to verify that the acoustic diffusivity $D_a$ agrees with the (relativistic) Navier-Stokes prediction with the transport coefficients given in \cite{Weinberg1971,UdeyIsrael1982}. It can be easily verified that, when $\kappa_p=0$, the acoustic diffusivity predicted by Navier-Stokes reads\footnote{While the first two terms in the numerator of \eqref{daa} are well-known, the third term is usually neglected. To see where it comes from, consider that the perturbation to the momentum density is $\delta T^{01}=(\rho{+}P)\delta u^1 {+} \delta q^1$, where $\delta q^1 {=} {-}\kappa (\partial_1 \delta T{+}T\partial_t\delta u^1)$ is the heat flux \cite{Hiscock_Insatibility_first_order}. Since, in our case, $\delta T \approx 0$ and $\partial_t^2 \approx c_s^2 \partial^2_x$, we have that $\partial_t \delta T^{01} \approx(\rho {+}P)\partial_t \delta u^1-c_s^2 T \kappa \partial^2_x \delta u^1$, meaning that $c_s^2 T \kappa$ can be affectively added to the bulk viscosity. Clearly, this is a purely relativistic effect. For more detials, see \cite[Eq.s (2.55) and (2.57)]{Weinberg1971}}
\begin{equation}\label{daa}
D_a= \dfrac{\frac{4}{3} \eta{+}\zeta{+}Tc_s^2 \kappa }{2(\rho {+}P)} \, ,
\end{equation}
where $\zeta$ is the bulk viscosity coefficient. Comparing \eqref{daa} with \eqref{cstotdaa}, and invoking \eqref{etuzzo} and \eqref{kappuzzo}, we find that
\begin{equation}\label{zetuzzo}
\zeta=\dfrac{4\pi^2}{135} T^4 \tau= \dfrac{4}{9} aT^4 \tau \, ,
\end{equation}
which agrees with the formula of \cite{Weinberg1971,UdeyIsrael1982} since, in our fluid of interest,
\begin{equation}
    \dfrac{\partial P_\text{tot}}{\partial \rho_\text{tot}} \bigg|_n= \dfrac{\nu}{3(1{+}\nu)} \xrightarrow[]{\nu \rightarrow 0} 0 \, ,
\end{equation}
see equation \eqref{grienz}. In conclusion, we confirm that, in the opticallly thick limit, radiation hydrodynamics reduces to relativistic Navier-Stokes, and its transport coefficients $\{\eta,\kappa,\zeta \}$ are indeed those provided by \citet{Weinberg1971}.

\subsection{Optically thin limit of shear waves}

If we take the limit of \eqref{mainformula} as $k\tau \rightarrow \infty$, we obtain
\begin{equation}\label{frequencyklarge}
\omega(k) \longrightarrow -5iD_s/\tau^2 \, .
\end{equation}
We can explain this asymptotic behavior with the following simple model. Consider a periodic rectangular shear wave, where layers with velocity $\delta u^3$ alternate with layers with velocity $-\delta u^3$. Suppose that at $t=0$ the photons are in thermal equilibrium with the medium (inside each layer). Then, suppose that they are released, and they travel at the speed of light till time $t \approx \tau$, when they are absorbed by the medium. If $k\tau$ is large, the layers with alternating velocity are very thin, compared to the distance traveled by the photons. Hence, the photons cross many layers before being absorbed, and they roughly have $50$ \% probability of being absorbed by a layer that moves with velocity $\delta u^3$ and $50$ \% probability of being absorbed by a layer that moves with velocity $-\delta u^3$. This means that the part of the fluid that moves with velocity $\delta u^3$ loses half of its photons, and it receives half of the photons that were belonging to the part of the fluid that moves with velocity $-\delta u^3$. This leads to the following change in momentum density:
\begin{equation}\label{mommuto}
T^{03}(\tau){-}T^{03}(0) =- \binom{\text{Momentum of}}{\text{photons lost}} + \binom{\text{Momentum of}}{\text{photons gained}} = -\dfrac{1}{2} (+T_R^{03}) + \dfrac{1}{2} (-T_R^{03}) = -T_R^{03} \, .
\end{equation}
Recalling that the parameter \eqref{lambo} is small, we have $T^{03} \approx T_M^{03}= (\rho{+}P) \delta u^3$. Since the photons were initially in thermal equilibrium with the medium, we have $\delta T_R^{03}=4aT^4 \delta u^3/3$. Thus, dividing both sides of \eqref{mommuto} by $\tau$, we obtain
\begin{equation}\label{belluto}
\partial_t \delta u^3=- 5 \dfrac{D_s}{\tau^2} \delta u^3 \, ,
\end{equation}
which results precisely in the relaxation frequency \eqref{frequencyklarge}. Equation \eqref{belluto} may be viewed as the ``shear-wave analog'' of Newton's law of cooling \cite[\S 100]{mihalas_book}.

\subsection{Optically thin limit of sound waves}

If we take the limit of \eqref{mainformula3} as $k\tau \rightarrow \infty$, we obtain
\begin{equation}
\omega(k) \longrightarrow c_s k -5 i D_s/(2\tau^2) \, .
\end{equation}
To have an intuitive understanding of this behavior, we can invoke a similar model to the one we used for shear waves, just replacing $\delta u^3$ with $\delta u^1$. Then, the calculation of the momentum exchange due to photons is the same as in the previous subsection [equations \eqref{mommuto} and \eqref{belluto}]. The only difference is that, now, the wave is longitudinal, and a change in $T^{01}$ can also be caused by pressure gradients. Hence, equation \eqref{belluto} is now replaced by the following system:
\begin{equation}\label{bellutowow}
\begin{split}
& \dfrac{\partial_t \delta P}{\rho {+} P} +c_s^2 \partial_1 \delta u^1 =0 \\
& \partial_t \delta u^1+\dfrac{\partial_1 \delta P}{\rho{+}P}=- 5 \dfrac{D_s}{\tau^2} \delta u^1 \, , \\
\end{split}
\end{equation}
where the first line is the continuity equation of baryons (with $\kappa_p =0$). Combining these two equations, we obtain a telegraph-type equation for the velocity:
\begin{equation}
\partial^2_t \delta u^1+5\dfrac{D_s}{\tau^2} \partial_t \delta u^1 =c_s^2 \partial^2_1 \delta u^1 \, .
\end{equation}
The corresponding dispersion relations can be computed analytically. Recalling that we are working in the limit of very large $k\tau$, we obtain the desired dispersion relation,
\begin{equation}
\omega(k)=\pm c_s k -i \dfrac{5D_s}{2\tau^2} \, ,
\end{equation}
which is exactly what we were looking for. Note that, while in the optically thick limit the group velocity of \eqref{mainformula3} is the \textit{combined} speed of sound $c_s^{\text{tot}}$ of matter+radiation, in the optically thin limit it is the speed of sound $c_s$ of matter alone. This is because, at small $k\tau$, matter and radiation are tightly coupled and oscillate together, while, at large $k\tau$, radiation effectively decouples and just spreads around uniformly, so that only matter oscillates.

\subsection{Optically thin limit of heat waves}

If we take the limit of \eqref{mainformula2} as $k\tau \rightarrow \infty$, we get
\begin{equation}\label{frequencyklarge2}
\omega(k) \longrightarrow -3iD_h/\tau^2 \, .
\end{equation}
Also here, there is a simple explanation. Consider a periodic rectangular heat wave, where layers with temperature perturbation $\delta T$ alternate with layers with temperature perturbation $-\delta T$. For simplicity, we set $\kappa_p = 0$, so thermal expansion can be neglected. At $t=0$, the photons are in local equilibrium with the fluid. As before, they then travel at the speed of light till $t \approx \tau$, when they are absorbed. Again, if $k\tau$ is very large, the layers with alternating temperature are very thin and the photons roughly have $50$ \% probability of being absorbed by a layer with temperature $T+\delta T$ and $50$ \% probability of being absorbed by a layer with temperature $T{-}\delta T$. This means that the part of the fluid that has temperature $T{+}\delta T$ loses half of its photons, and receives half of the photons coming from the part of the fluid with temperature $T{-}\delta T$. The resulting change in entropy density is (entropy is conserved in the linear regime):
\begin{equation}\label{mommuto2}
\delta s^{0}(\tau){-}\delta s^{0}(0) =- \binom{\text{Entropy of}}{\text{photons lost}} + \binom{\text{Entropy of}}{\text{photons gained}} = -\dfrac{1}{2} (s_R^0+\delta s_R^{0}) + \dfrac{1}{2} (s_R^0-\delta s_R^{0}) = -\delta s_R^{0} \, .
\end{equation}
Recalling that the parameter \eqref{fourtieight} is small, we have $s^{0} \approx s_M^{0}=n \mathfrak{s}$. Since the photons were initially in thermal equilibrium with the medium, we have $\delta s_R^{0}=\delta (4aT^3/3)=4aT^2 \delta T$. Thus, dividing both sides of \eqref{mommuto2} by $\tau$, we obtain
\begin{equation}\label{Newton}
\partial_t \delta \mathfrak{s}=- 3 D_h \delta \mathfrak{s}/\tau^2 \, ,
\end{equation}
which results precisely in the relaxation frequency \eqref{frequencyklarge2}. Equation \eqref{Newton} is just Newton's law of cooling \cite[\S 100]{mihalas_book}.

\vspace{-0.2cm}
\section{Mathematical discussion (diffusive modes only)}
\vspace{-0.2cm}

\subsection{Covariant stability of shear and heat waves}

\begin{figure}[b!]
\begin{center}
\includegraphics[width=0.46\textwidth]{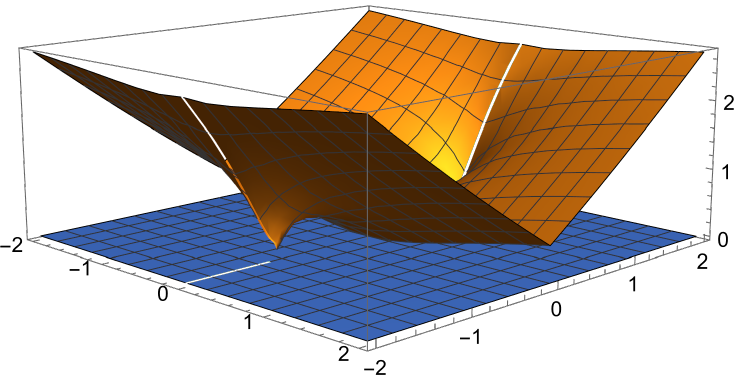}
\includegraphics[width=0.46\textwidth]{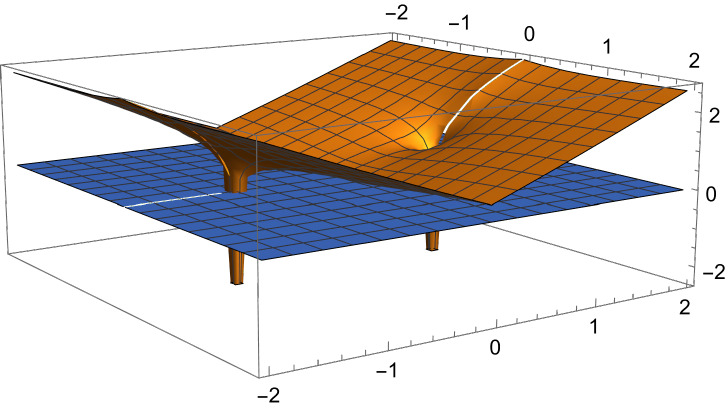}
	\caption{Graph of the function $G(\q_R,\q_I)$ according to equation \eqref{mainformulacheck} (left panel) and \eqref{mainformulacheck2} (right panel), for $\lambda{=}2.5$ and $\nu{=}0.3$. The blue plane marks level $0$. If the orange surface deeps below the plane, the dispersion relation is not covariantly stable.}
	\label{fig:covstab}
	\end{center}
\end{figure}

A dispersion relation $\omega(k) {:}\,\mathbb{C}\,{\rightarrow}\, \mathbb{C}$ is said to be ``covariantly stable'' if it cannot be Lorentz-transformed into a growing Fourier mode \cite{Hiscock_Insatibility_first_order,GavassinoSuperluminal2021}. It was proven that $\omega(k)$ is covariantly stable if and only if $\mathfrak{Im} \, \omega(k) \leq |\mathfrak{Im} \, k |$ for all $k$ complex \cite{GavassinoBounds2023}. This is equivalent to requiring that the function $G(\q) = |\mathfrak{Im} \, \q|-\mathfrak{Re} \, \Gamma(\q)$ be non-negative for all choices of $\q \in \mathbb{C}$. If we write $\q=\q_R {+}i \q_I$, with $\q_R,\q_I \in \mathbb{R}$, the quantity $G$ is a function from $\mathbb{R}^2$ to $\mathbb{R}$, whose explicit form is
\begin{flalign}\label{mainformulacheck}
\text{Shear waves:} \quad \quad G (\q_R,\q_I) = |\q_I|+\lambda \, \mathfrak{Re}\bigg[ \dfrac{2}{3}+\dfrac{1}{(\q_R{+}i\q_I)^2}-\bigg( 1{+}\dfrac{1}{(\q_R{+}i\q_I)^2} \bigg) \dfrac{\arctan(\q_R {+}i \q_I)}{\q_R {+}i\q_I}  \bigg] \, ; &&
\end{flalign}
\begin{flalign}\label{mainformulacheck2}
\text{Heat waves:} \quad \quad \, \, G (\q_R,\q_I) = |\q_I|+\nu \, \mathfrak{Re}\bigg[ 1- \dfrac{\arctan(\q_R {+}i \q_I)}{\q_R {+}i\q_I}  \bigg]\, . &&
\end{flalign}
See Figure \ref{fig:covstab} for the 3D plots of these two functions. It turns out that, for shear waves, $G$ is non-negative all the way to $\lambda \gtrsim 2.5$. Thus, the dispersion relation \eqref{mainformula} is covariantly stable also outside its formal regime of applicability. Instead, for heat waves, $G$ always becomes infinitely negative near $\q=\pm i$, meaning that \eqref{mainformula2} is not covariantly stable.

The fact that the dispersion relation for heat waves becomes unstable (in some boosted frame \cite{GavassinoBounds2023}) is a signal that, for $\q \approx \pm i$, our formal derivation of \eqref{mainformula2} breaks down. This is no surprise, since both \eqref{mainformula} and \eqref{mainformula2} were derived assuming that $\q$ was real. To understand the origin of the problem, consider again the integral in equation \eqref{8settatntuno}. If we set $\q= i$, and expand around $\Gamma=0$, the denominator $1{+}\Gamma {+} i \q \xi$ becomes $1- \xi$. When this happens, the integral diverges, since
\begin{equation}
\int_{-1}^1 \dfrac{d\xi}{1+\Gamma - \xi} = \ln(2{+}\Gamma)-\ln \Gamma  \xrightarrow[]{\Gamma \longrightarrow 0} +\infty \, .
\end{equation}
It follows that, when we look for an approximate solution of the system \eqref{8settatntuno}-\eqref{8settantcinque} with $\q=\pm i$, the equations $F(\Gamma,\delta P,\delta u^1,\nu){=}0$ are irregular at $(0,0,0,0)$, and the assumptions of the implicit function theorem do not hold. 

Acknowledging that the dispersion relation \eqref{mainformula2} of heat waves is not covariantly stable should not prevent us from using it. In fact, equation \eqref{mainformula2} remains a good approximation of kinetic theory whenever $\delta \mathfrak{s}$ is a superposition of modes with $k \in \mathbb{R}$. More precisely: Equation \eqref{mainformula2} can be used to compute $\delta \mathfrak{s}(t,x)$ for $t>0$ whenever $\delta \mathfrak{s}(0,x)$ has a well-defined Fourier transform. Note that, since $-3D_h/\tau^2 \leq \mathfrak{Im} \omega \leq 0$, the spatial profile of $\delta \mathfrak{s}$ has a well-defined Fourier transform at $t>0$ if and only if it has a well-defined Fourier transform at $t=0$.

\subsection{Causality considerations}

It was recently shown \cite{GavassinoDispersion2024} that, in a dispersive (stable) system, there is no way to assign a notion of causality to a \textit{single} dispersion relation $\omega_0(k)$. In fact, it was proven in \cite{GavassinoDispersion2024} that (independently from the system's details) the collective excitation $\delta\Psi_0(t,x)$ that propagates according to $\omega_0(k)$ can never be localized, so it is not possible to define a speed of propagation. For example, suppose that the collective excitation of interest is a heat wave, and we have constructed an initial state such that the temperature perturbation $\delta T(0,x)$ is contained within a finite region of space. Then, the perturbation $\delta Y(0,x)$ to some other quantity $Y$ (e.g., the heat flux) \textit{must} cover the whole space. 

\begin{figure}[b!]
\begin{center}
\includegraphics[width=0.44\textwidth]{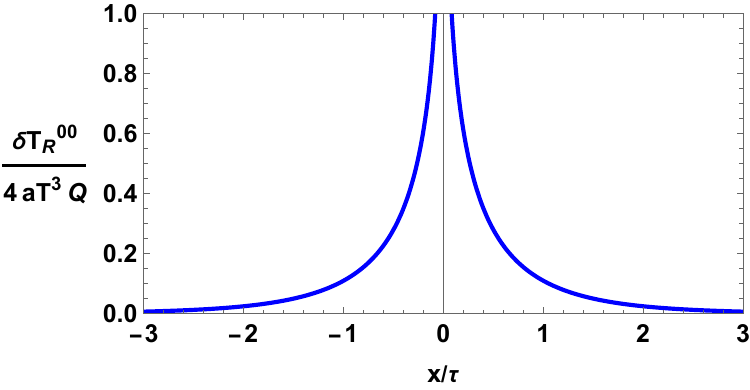}
	\caption{Radiation energy density associated to a heat wave (with dispersion relation \eqref{mainformula2}) whose temperature fluctuation is a Dirac delta centered in the origin, namely $\delta T(x) =Q\delta(x/\tau)$. Note that the divergence of $\delta T_R^{00}(x)$ in the origin is logarithmic.}
	\label{fig:Ei}
	\end{center}
\end{figure}

Let us confirm that this general result applies also to radiation hydrodynamics. We take, as our collective excitation of interest, the heat waves, with dispersion relation \eqref{mainformula2} (shear and sound waves are analogous). We assume that the initial temperature perturbation $\delta T(0,x)$ has compact support, and we study the initial perturbation to the radiation energy density, $\delta T^{00}_R(0,x)$. Using \eqref{stressOne} and \eqref{df}, we find that, to first order in $\nu$,
\begin{equation}\label{relazionarriones}
\dfrac{\delta T_R^{00}(0,x)}{4aT^3} = \int \dfrac{d \q}{2\pi} e^{i\q x/\tau}\, \delta T(\q) \, \dfrac{\arctan(\q)}{\q}\, .
\end{equation}
Now, since $\delta T(0,x)$ is compactly supported, its Fourier transform $\delta T(\q)$ is an entire function of $\q {\in} \mathbb{C}$ \cite[Th 7.1.14]{HormanderOperatorsBook}. Hence, the Fourier transform of $\delta T_R^{00}(0,x)$ is the product of an entire function with the function $\arctan(\q)/\q$, which has two branch cuts, starting at $\q=\pm i$. It follows that the Fourier transform of $\delta T_R^{00}(0,x)$ cannot be entire, meaning that the support of $\delta T_R^{00}(0,x)$ is unbounded (again by \cite[Th 7.1.14]{HormanderOperatorsBook}). For example, if $\delta T(0,x)\propto \delta (x)$, then $\delta T_R^{00}(0,x)\propto -\text{Ei}(-|x/\tau|)/2$, where $\text{Ei}$ is the exponential integral Ei, which has infinite support, see figure \ref{fig:Ei}.

In summary: It is impossible to simultaneously localize the perturbations to the fluid temperature $T$ and to the radiation energy density $T_R^{00}$ without turning on some additional excitation mode that does not follow \eqref{mainformula2}, like, e.g. a non-hydrodynamic mode \cite{GavassinoGapless2024rck}. If all collective excitations that do not follow \eqref{mainformula2} are set to zero, the propagation of $\delta T(t,x)$ is governed by seemingly non-local dynamics, because the relationship \eqref{relazionarriones} between $\delta T$ and $\delta T_R^{00}$ is non-local. Such non-locality does not violate relativistic causality, because \eqref{relazionarriones} describes a correlation, not a direct causation.

\subsection{Evolution of jump-discontinuities}
\vspace{-0.2cm}

The dynamics of jump-discontinuities and wavefronts depends on the asymptotic behaviour of the dispersion relations $\omega(k)$ in the UV limit (i.e. at large $k$). In fact, jump-discontinuities of the fluid variables manifest themselves, in Fourier space, as infinitely long tails (which decay like $\sim 1/k$), and such tails are multiplied by a dynamical factor $e^{-i\omega(k)t}$ that determines the time-dependence of the jump structure. For example, the ordinary diffusion equation ($\omega {\sim} -ik^2$) suppresses all power-law tails at any $t>0$, since it multiplies them by a Gaussian factor $\sim e^{-k^2 t}$. As a result, discontinuities and wedges are immediately smoothed out at positive times \cite[\S 2.3.3]{EvansPDEsBook}. By contrast, Cattaneo's theory of diffusion \cite{cattaneo1958} multiplies the UV tails by a factor $\sim e^{(iak-b)t}$, so jump-discontinuities travel with speed $a\sim \sqrt{D/\tau} \neq 0$ (called ``second sound speed'' \cite{rezzolla_book}), and their magnitude decays exponentially at rate $b$. 

\begin{figure}[b!]
\begin{center}
\includegraphics[width=0.48\textwidth]{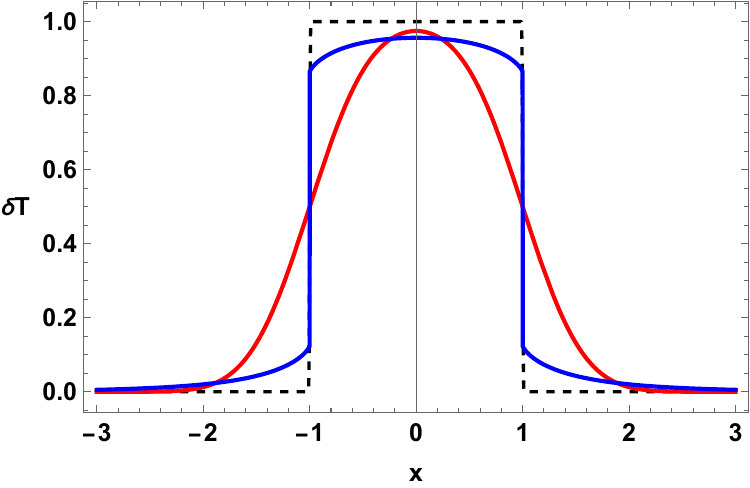}
\includegraphics[width=0.48\textwidth]{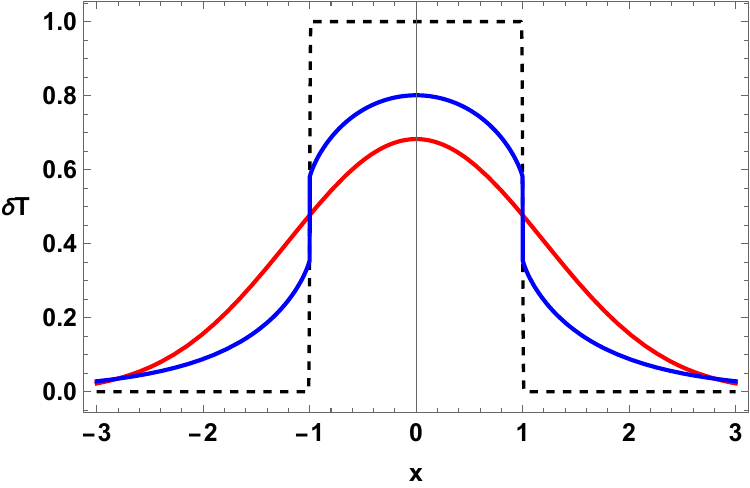}
\includegraphics[width=0.48\textwidth]{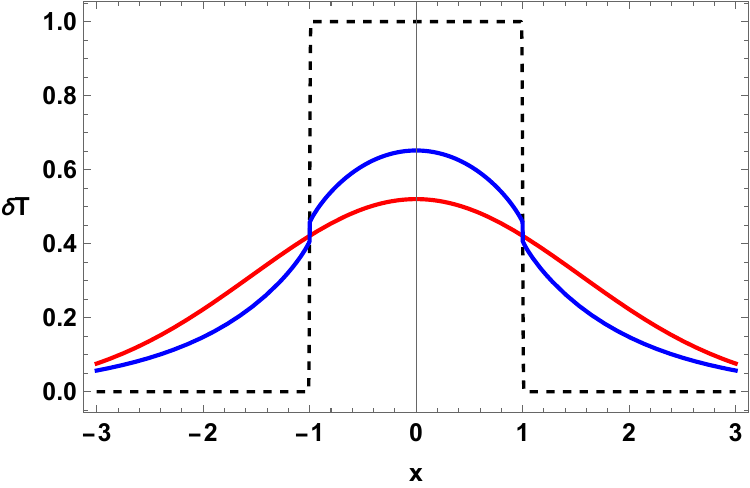}
\includegraphics[width=0.48\textwidth]{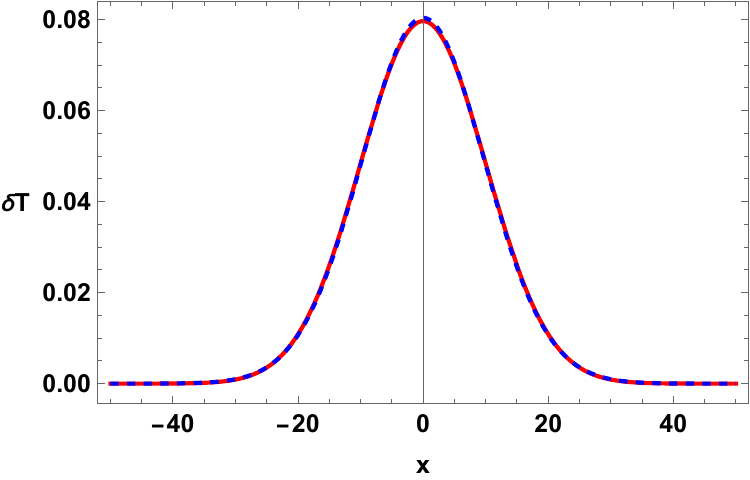}	\caption{Evolution of the discontinuous temperature profile \eqref{taintanone} with dispersion relation \eqref{mainformula2} (blue) compared with ordinary diffusion, $\omega(k)=-iD_h k^2$ (red). The initial data (dashed black) is $\delta T(0,x)=\Theta(1-x^2)$, and we work in spacetime units so that $\tau=1$. Each panel is a snapshot at a different time, respectivly $D_ht=0.1$ (up-left), $0.5$ (up-right), $1$ (down-left), $50$ (down-right). Note that the down-right panel differs from the others in that (a) the range of both axes is different, (b) we are no longer plotting the initial data, and (c) the blue curve is now dashed, since nearly overlaps the red one.}
	\label{fig:Comparison}
	\end{center}
\end{figure}

Interestingly, the analytical dispersion relations \eqref{mainformula} and \eqref{mainformula2} of diffusive modes in coupled radiation-matter systems exhibit a different behaviour from both ordinary diffusion and Cattaeo diffusion. In fact, at large $k$, the frequency is just $\omega \sim -ib$, meaning that jump-discontinuities stand still at their initial location (no second sound), and their amplitude decays exponentially by a factor $\sim e^{-bt}$. This behavior is well-illustrated by the numerical example in figure \ref{fig:Comparison}, where we compare a solution of the usual diffusion equation ($\partial_t \delta T=D_h \partial^2_x \delta T$) with initial data $\delta T(0,x)=\Theta(1-x^2)$ and the corresponding solution of radiation hydrodynamics,
\begin{equation}\label{taintanone}
    \delta T(t,x)= \int_\mathbb{R}  \dfrac{\sin(k)}{\pi k} e^{ikx-i\omega(k) t} dk \, ,
\end{equation}
computed using the dispersion relation \eqref{mainformula2}. As can be seen, discontinuities evolve quite differently. The interested reader can see \cite{GavassinoShearRadiation2024jej} for a similar calculation with shear waves.

\section{Comparison with M1 closure predictions}

A widely used approximation in radiation-hydrodynamic simulations is the M1 closure scheme \cite{Minerbo1978,Levermore1984,Sadowski2013,Fragile:2014bfa,Fragile:2018xee,Murchikova:2017zsy,Anninos2020,GavassinoRadiazione}. This approach only tracks the first two moments of the radiation distribution function, namely $\{\varepsilon_R(x^\alpha),F^\nu(x^\alpha)\}$, which represent respectively the radiation energy density and the radiation energy flux in the local rest frame of the medium  (with $F^\nu u_\nu =0$). The radiation stress-energy tensor, then, is approximated as follows: 
\begin{equation}\label{dedkorfc}
T_R^{\mu \nu}= \dfrac{4}{3} \varepsilon_R u^\mu u^\nu +\dfrac{1}{3}\varepsilon_R g^{\mu \nu } +F^\mu u^\nu + u^\mu F^\nu+ \dfrac{3\chi{-}1}{2} \varepsilon_R \bigg[ \dfrac{F^\mu F^\nu}{F^\alpha F_\alpha} {-}\dfrac{g^{\mu \nu} {+}u^\mu u^\nu}{3} \bigg] \, , 
\end{equation}
where $\chi$ is the Eddington factor \cite{Minerbo1978,Levermore1984}, which is assumed to be a function of the scalar $F^\alpha F_\alpha/\varepsilon_R^2$. The last term in \eqref{dedkorfc} is the closure, since it expresses the radiative stress tensor as a function of $\{\varepsilon_R(x^\alpha),F^\nu(x^\alpha)\}$. To derive an equation of motion for $\varepsilon_R$ and $F^\nu$, one can just combine \eqref{Boltzmann} with \eqref{stressOne}, which results in an \textit{exact} balance law:
\begin{equation}\label{basilisco}
\partial_\mu T_R^{\mu \nu}= -\dfrac{1}{\tau} (\varepsilon_R {-}aT^4) u^\nu -\dfrac{1}{\tau} F^\nu \, .
\end{equation}
Let us compare the hydrodynamic dispersion relations of this approximate ``fluid-type'' model of radiation with \eqref{mainformula}-\eqref{mainformula3}.

\subsection{Linearised radiation-hydrodynamic equations with M1 closure}

Let us linearize equation \eqref{basilisco}, together with the usual conservation laws of energy, momentum, and baryons. In all approaches of interest, one assumes that, for small $F^\nu$, the Eddington factor can be expanded as $\chi \approx 1/3 +z \, F^\alpha F_\alpha/\varepsilon_R^2$, for some number $z$. Hence, in the linear regime, the pressure anisotropy in \eqref{dedkorfc} vanishes, and the M1 closure reduces to the Eddington approximation. Introducing the notation  
$\delta \mathcal{E}=\delta \varepsilon_R/\varepsilon_R$ and $\delta \mathcal{F}^j =\delta F^j/\varepsilon_R +4\delta u^j/3$, we obtain the following system:
\begin{flalign}\label{Asettatntuno}
\boxed{\rho^{\textcolor{white}{1}}} \quad  \omega \, \delta \mathfrak{s} = \dfrac{aT^3}{n} \bigg[  k \, \delta \mathcal{F}^1 - \omega \, \delta \mathcal{E} \bigg]  \, , &&
\end{flalign}
\begin{flalign}\label{Asettantadue}
\boxed{u^1} \quad \omega \,  \delta u^1-k\dfrac{ \delta P}{\rho{+}P} =\dfrac{aT^4}{\rho{+}P} \bigg[k \dfrac{1}{3}  \delta \mathcal{E}- \omega \, \delta \mathcal{F}^1    \bigg] \, , &&
\end{flalign}
\begin{flalign}\label{Asettantatre}
\boxed{u^j} \quad  \omega \, \delta u^j=-\omega \dfrac{aT^4}{\rho{+}P}  \delta \mathcal{F}^j   \quad \spc \spc \spc \spc  \spc (\text{for } j=2,3) \, , &&
\end{flalign}
\begin{flalign}\label{Asettantcinque}
\boxed{n^{\textcolor{white}{1}}} \quad   \delta u^1 = \dfrac{\omega}{k} \dfrac{\delta n}{n} \, , &&
\end{flalign}
\begin{flalign}\label{Asettantasei}
\boxed{\varepsilon^{\textcolor{white}{1}}} \quad (1{-}i\omega \tau)\delta \mathcal{E}=4\dfrac{\delta T}{T}-ik\tau  \delta \mathcal{F}^1  \, , &&
\end{flalign}
\begin{flalign}\label{Asettantasette}
\boxed{F^1} \quad (1{-}i\omega \tau)\delta \mathcal{F}^1=-\dfrac{1}{3} ik \tau  \delta \mathcal{E}+\dfrac{4}{3}\delta u^1  \, , &&
\end{flalign}
\begin{flalign}\label{Asettantotto}
\boxed{F^j} \quad (1{-}i\omega \tau)\delta \mathcal{F}^j= \dfrac{4}{3} \delta u^j \spc \spc \spc \spc \spc \quad (\text{for } j=2,3) \, , &&
\end{flalign}
where, as usual, we have assumed that all perturbed fields have a spacetime dependence of the form $e^{ikx^1-i\omega t}$.

\subsection{Shear waves}

In \cite{AndersonSpiegel1972,GavassinoRadiazione,GavassinoFronntiers2021}, it was argued that the optically thick limit of a radiation-hydrodynamic system with M1 closure is a viscous fluid, with the same values of $\zeta$ and $\kappa$ as in \cite{Weinberg1971}, but with $\eta=0$. Hence, the damping of shear waves cannot be correctly described by M1 models. Indeed, it was shown in \cite{GavassinoShearRadiation2024jej} that shear waves with M1 closure do not decay. This can be seen directly from the system \eqref{Asettatntuno}-\eqref{Asettantotto}. The two pairs of degrees of freedom $\{\delta u^2,\delta \mathcal{F}^2 \}$ and $\{\delta u^3,\delta \mathcal{F}^3 \}$ fully decouple from all other degrees of freedom, and their fluctuations are shear modes, governed by equations \eqref{Asettantatre} and \eqref{Asettantotto}. We find that the state $\delta \mathcal{F}^3(k)=4\delta u^3(k)/3$ solves equations \eqref{Asettantatre} and \eqref{Asettantotto} with $\omega(k)=0$ for all $k$, meaning that M1 fluids possess shear wave solutions that survive forever, in sharp contrast with \eqref{mainformula}.

\subsection{Heat waves}

The M1 closure scheme is known to describe heat propagation quite accurately, both in the optically thick and in the optically thin limit \cite{Sadowski2013}. Indeed, the dispersion relation \eqref{mainformula2} and its M1 analogue are textbook material \cite[\S 100]{mihalas_book}. Let us briefly summarize the result.

The derivation of the dispersion relation in M1 systems is analogous to that in section \ref{heattoneee}, namely set $\delta u^2=\delta u^3=0$, and $\delta \mathfrak{s}=1$. Introduce the small parameter $\nu$ defined in equation \eqref{nuino}, and expand all variables to first order in $\nu$. We do not report the intermediate steps (which are not so enlightening) and we just provide the final formula:  
\begin{figure}
\begin{center}
\includegraphics[width=0.51\textwidth]{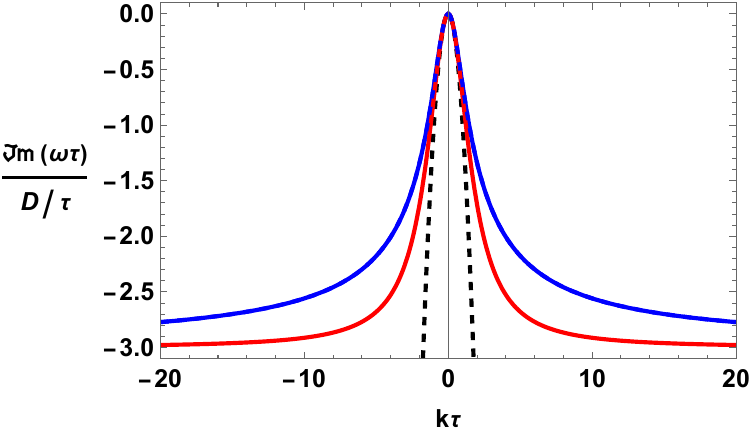}
\includegraphics[width=0.45\textwidth]{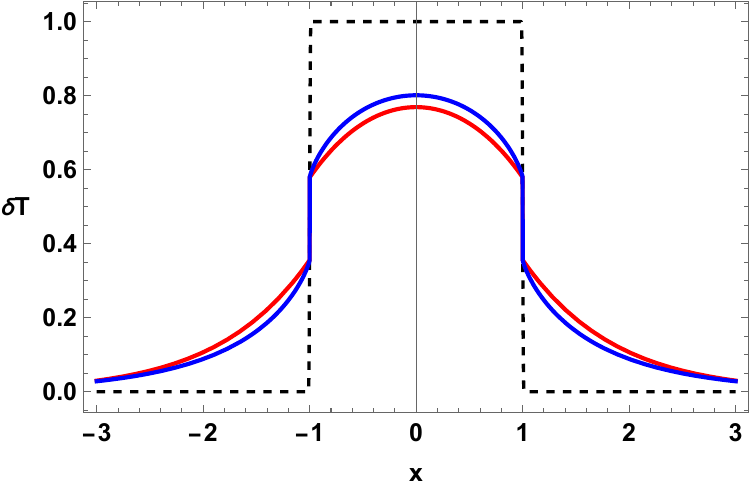}
	\caption{Comparison between the dispersion relation \eqref{mainformula2} computed directly from the radiative transport equation (blue) and the dispersion relation \eqref{m1one} computed assuming M1 closure (red). Left panel: Imaginary part of the dispersion relations, where the reference dashed line is ordinary diffusion, i.e. $\omega=-iD_h k^2$ (the real part vanishes in all models). Right panel: Temperature profile $\delta T(t,x)$ at time $t=\tau^2/(2D_h)$, with initial data $\delta T(0,x)=\Theta(1-x^2)$ (dashed line). The Fourier integral representing both solutions is reported in equation \eqref{taintanone}. As in figure \ref{fig:Comparison}, the spacetime units are chosen such that $\tau=1$.}
	\label{fig:M1}
	\end{center}
\end{figure}
\begin{equation}\label{m1one}
\omega=-i \dfrac{D_h k^2}{1{+}(k\tau)^2/3} + \mathcal{O}\big(D_h^2/\tau^2\big)\, ,
\end{equation}
where $D_h$ is the same diffusivity coefficient appearing in \eqref{mainformula2}, whose value is provided in equation \eqref{fourtieight}. The behavior of the M1 model in the optically thick limit ($k\tau {\rightarrow} 0$) is consistent with the radiative transport equation. Also the optically thin ($k\tau {\rightarrow} \infty$) limiting behavior is accurate, since $\omega \rightarrow -3iD_h/\tau^2$, in full agreement with equation \eqref{frequencyklarge2}. This implies that discontinuities shrink with the correct relaxation rate, as can be seen from figure \ref{fig:M1}, right panel. The discrepancy between \eqref{mainformula2} and \eqref{m1one} is only relevant in intermediate regimes, where the M1 closure overestimates the damping rate (and therefore the diffusive nature) of heat waves, see figure \ref{fig:M1}, left panel.

\subsection{Sound waves}

The dispersion relation of sound waves in M1 fluids can be computed from \eqref{Asettatntuno}-\eqref{Asettantotto} following the same procedure as in section \ref{sounduzzonzone}. This involves setting $0=\delta u^2=\delta u^3=\delta P-1$, and carrying out a perturbative expansion in the parameter $\lambda$, defined in equation \eqref{lambo}. Assuming that $\kappa_p=0$, one obtains the following formula (see figure \ref{fig:M1sound}):
\begin{figure}
\begin{center}
\includegraphics[width=0.48\textwidth]{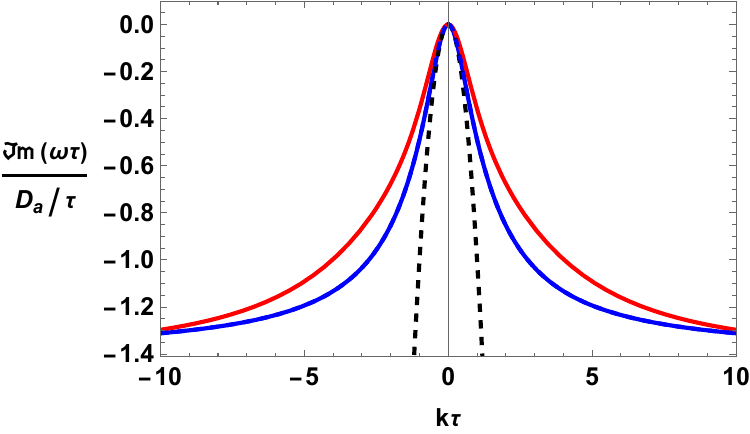}
\includegraphics[width=0.45\textwidth]{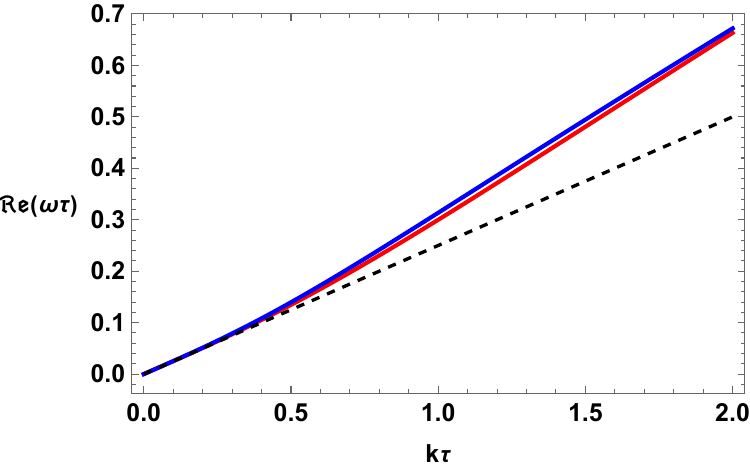}
	\caption{Comparison between the dispersion relation \eqref{mainformula3} computed directly from the radiative transport equation (blue) and the dispersion relation \eqref{muu1one} computed assuming M1 closure (red). The black dashed line is the Navier-Stokes limit $\omega=c_s^{\text{tot}}k-iD_a k^2$, whose coefficients are given in \eqref{cstotdaa}. We have made the following choices: $c_s=1/3$, and $D_s/\tau =1/10$.}
	\label{fig:M1sound}
	\end{center}
\end{figure}
\begin{equation}\label{muu1one}
\omega=c_s k -i\dfrac{15D_s}{2\tau^2} \bigg[ \dfrac{1}{3}-\dfrac{1{-}ic_s k\tau}{(k\tau)^2+3(1{-}ic_s k\tau)^2} \bigg] +\mathcal{O}(D_s^2/\tau^2)\, ,
\end{equation}
whose optically thin limit agrees with \eqref{mainformula3}.
For optically thick waves, we have $\omega(k)=c^{\text{tot}}_{M1} k{-}iD_{M1}k^2{+}\mathcal{O}(k^3 \tau^3)$, with
\begin{equation}\label{csdaaM1}
\begin{split}
c^{\text{tot}}_{M1}={}& \bigg(1-\dfrac{5D_s}{2\tau} \bigg) c_s \, , \\
D_{M1}={}& \dfrac{5}{6}(1+3c_s^2)D_s  \, .\\
\end{split}
\end{equation}
Comparing \eqref{csdaaM1} with \eqref{cstotdaa}, we see that the M1 closure gives the correct conglomerate speed of sound (i.e. $c^\text{tot}_{M1}=c^\text{tot}_s$), but not the correct acoustic diffusivity (i.e. $D_{M1} \neq D_a$).  This is expected, since M1 fluids have vanishing shear viscosity \cite{GavassinoRadiazione}. Indeed, one would obtain the correct value of $D_{M1}$ by simply setting $\eta=0$ in equation \eqref{daa}. 






\section{Discussion and Conclusions}
\vspace{-0.2cm}

We have solved the equations of relativistic radiation hydrodynamics in the linear regime,
under the assumptions \textbf{(a)}, \textbf{(b)}, and \textbf{(c)} listed in the introduction. Both matter and radiation have been evolved self-consistently, the former being subject to fluid-dynamical conservation laws, and the latter being governed by the relativistic Boltzmann equation. This led us to the dispersion relations \eqref{mainformula}, \eqref{mainformula2}, and \eqref{mainformula3}, which are exact to first order in the dimensionless parameters $D_s/\tau$ and $D_h/\tau$, whose magnitude scales like that of the ratios $T^{\mu \mu}_R/T^{\mu \mu}_M$.

The dispersion relation of shear waves agrees with our recent calculation in \cite{GavassinoShearRadiation2024jej}. The dispersion relation of heat waves agrees with that of \citet{Spiegel1957}, with the specific heat $c_p$ in place of $c_v$ (just as in ordinary fluid mechanics \cite{landau6}). To the best of our knowledge, the formula for the dispersion relation of sound waves in special relativity is completely new, although it is derived under the additional assumption that the matter component does not expand as it absorbs heat (see footnote \ref{kpvanishes}). Despite this limitation in the thermal properties of the matter sector, the radiation sector is evolved exactly, and all radiative corrections to sound propagation (e.g. radiation pressure, radiative viscosity, radiative heat transport, and acceleration-driven heat) are accurately captured, also at relativistic sound-speeds.

What do we learn from equations \eqref{mainformula}-\eqref{mainformula3}? In our opinion, the most important insights come from the observation that all three dispersion relations were computed directly from the exact linearized radiative transport equations \eqref{settatntuno}-\eqref{settantasei} alone, with the aid of perturbation theory techniques. This allows us to apply the ``machinery'' of theoretical relativistic fluid mechanics \cite{MooreCuts2018mma,Romatschke2017,FlorkowskiReview2018,RochaReview2023ilf}, and draw the following conclusions about radiation hydrodynamics as a whole:
\begin{itemize}
\item The dispersion relations \eqref{mainformula}-\eqref{mainformula3} are the hydrodynamic poles of the retarded linear response Green function of photon kinetic theory. Hence, if we expand them in Taylor series for small wavenumbers, i.e. $\omega(k){=}\sum_n a_n k^n$, there is a one-to-one correspondence between the Taylor coefficients $a_n$ and the infinite list of Chapman-Enskog transport coefficients of linearised radiation hydrodynamics \cite{McLennan1965,Dudynski1989,Struchtrup2011,HellerHydroHedron2023jtd}. In other words, the full knowledge of the Chapman-Enskog expansion (up to infinite order) is contained within \eqref{mainformula}-\eqref{mainformula3}. 
\item The radius of convergence of the Taylor series $\omega(k)=\sum_n a_n k^n$ is simply $\tau^{-1}$ for the diffusive modes, while it is $\tau^{-1}/(1{+}c_s)$ for the sound modes. This implies that the Chapman-Enskog expansion is well-defined (in rigorous mathematical terms \cite{McLennan1965,Dudynski1989}) in the linear regime \cite{GavassinoChapmanEnskog2024xwf}. Furthermore, the radiation mean free path $\tau$ marks the breakdown scale of the viscous hydrodynamic approximation.
\item Given that the Taylor series of $\omega(k)$ has a finite radius of convergence, the dispersion relations \eqref{mainformula}-\eqref{mainformula3} propagate matter waves faster than light \cite{GavassinoDispersion2024}. However, this does not entail superluminal signaling. In fact, if the initial wave profile at $t=0$ is built as a superposition of hydrodynamic excitations that obey \eqref{mainformula}-\eqref{mainformula3}, then the initial radiation field is not compactly supported. Hence, some radiation ``forerunners'' are visible to all observers already $t=0$, and the matter wave transports no new information. 
\item The transport coefficients $\eta$, $\kappa$, and $\zeta$ computed by \citet{Weinberg1971} coincide with those computed from our dispersion relations, and the latter coefficients are those that one would obtain from the respective Kubo formulas \cite{peliti_book,Czajka:2017bod}.
\item The sign of the term $(k\tau)^4 $ in equation \eqref{seriaccio} shows that the Super-Burnett approximation (i.e. third-order viscous hydrodynamics \cite{SuperBurnett1993}) is unstable in fluids with radiation.
\end{itemize}

\section*{Acknowledgements}

This work is partially supported by a Vanderbilt's Seeding Success Grant.

\bibliography{Biblio}

\begin{thebibliography}{61}%
\makeatletter
\providecommand \@ifxundefined [1]{%
 \@ifx{#1\undefined}
}%
\providecommand \@ifnum [1]{%
 \ifnum #1\expandafter \@firstoftwo
 \else \expandafter \@secondoftwo
 \fi
}%
\providecommand \@ifx [1]{%
 \ifx #1\expandafter \@firstoftwo
 \else \expandafter \@secondoftwo
 \fi
}%
\providecommand \natexlab [1]{#1}%
\providecommand \enquote  [1]{``#1''}%
\providecommand \bibnamefont  [1]{#1}%
\providecommand \bibfnamefont [1]{#1}%
\providecommand \citenamefont [1]{#1}%
\providecommand \href@noop [0]{\@secondoftwo}%
\providecommand \href [0]{\begingroup \@sanitize@url \@href}%
\providecommand \@href[1]{\@@startlink{#1}\@@href}%
\providecommand \@@href[1]{\endgroup#1\@@endlink}%
\providecommand \@sanitize@url [0]{\catcode `\\12\catcode `\$12\catcode `\&12\catcode `\#12\catcode `\^12\catcode `\_12\catcode `\%12\relax}%
\providecommand \@@startlink[1]{}%
\providecommand \@@endlink[0]{}%
\providecommand \url  [0]{\begingroup\@sanitize@url \@url }%
\providecommand \@url [1]{\endgroup\@href {#1}{\urlprefix }}%
\providecommand \urlprefix  [0]{URL }%
\providecommand \Eprint [0]{\href }%
\providecommand \doibase [0]{http://dx.doi.org/}%
\providecommand \selectlanguage [0]{\@gobble}%
\providecommand \bibinfo  [0]{\@secondoftwo}%
\providecommand \bibfield  [0]{\@secondoftwo}%
\providecommand \translation [1]{[#1]}%
\providecommand \BibitemOpen [0]{}%
\providecommand \bibitemStop [0]{}%
\providecommand \bibitemNoStop [0]{.\EOS\space}%
\providecommand \EOS [0]{\spacefactor3000\relax}%
\providecommand \BibitemShut  [1]{\csname bibitem#1\endcsname}%
\let\auto@bib@innerbib\@empty
\bibitem [{\citenamefont {{Prialnik}}(2009)}]{Prialnik2009}%
  \BibitemOpen
  \bibfield  {author} {\bibinfo {author} {\bibfnamefont {D.}~\bibnamefont {{Prialnik}}},\ }\href@noop {} {\emph {\bibinfo {title} {{An Introduction to the Theory of Stellar Structure and Evolution}}}}\ (\bibinfo  {publisher} {Cambridge University Press},\ \bibinfo {address} {Cambridge},\ \bibinfo {year} {2009})\BibitemShut {NoStop}%
\bibitem [{\citenamefont {{Pomraning}}(1973)}]{Pomraning1973}%
  \BibitemOpen
  \bibfield  {author} {\bibinfo {author} {\bibfnamefont {G.~C.}\ \bibnamefont {{Pomraning}}},\ }\href@noop {} {\emph {\bibinfo {title} {{The equations of radiation hydrodynamics}}}}\ (\bibinfo  {publisher} {Dover Publications, Mineola},\ \bibinfo {year} {1973})\BibitemShut {NoStop}%
\bibitem [{\citenamefont {{Mihalas}}\ and\ \citenamefont {{Weibel Mihalas}}(1984)}]{mihalas_book}%
  \BibitemOpen
  \bibfield  {author} {\bibinfo {author} {\bibfnamefont {D.}~\bibnamefont {{Mihalas}}}\ and\ \bibinfo {author} {\bibfnamefont {B.}~\bibnamefont {{Weibel Mihalas}}},\ }\href@noop {} {\emph {\bibinfo {title} {{Foundations of radiation hydrodynamics}}}}\ (\bibinfo  {publisher} {Oxford University Press, Oxford},\ \bibinfo {year} {1984})\BibitemShut {NoStop}%
\bibitem [{\citenamefont {Castor}(2004)}]{CastorRadiationBook_2004}%
  \BibitemOpen
  \bibfield  {author} {\bibinfo {author} {\bibfnamefont {J.~I.}\ \bibnamefont {Castor}},\ }\href@noop {} {\emph {\bibinfo {title} {Radiation Hydrodynamics}}}\ (\bibinfo  {publisher} {Cambridge University Press},\ \bibinfo {year} {2004})\BibitemShut {NoStop}%
\bibitem [{\citenamefont {{Thomas}}(1930)}]{Thomas1930}%
  \BibitemOpen
  \bibfield  {author} {\bibinfo {author} {\bibfnamefont {L.~H.}\ \bibnamefont {{Thomas}}},\ }\href {\doibase 10.1093/qmath/os-1.1.239} {\bibfield  {journal} {\bibinfo  {journal} {The Quarterly Journal of Mathematics}\ }\textbf {\bibinfo {volume} {1}},\ \bibinfo {pages} {239} (\bibinfo {year} {1930})}\BibitemShut {NoStop}%
\bibitem [{\citenamefont {{Weinberg}}(1971)}]{Weinberg1971}%
  \BibitemOpen
  \bibfield  {author} {\bibinfo {author} {\bibfnamefont {S.}~\bibnamefont {{Weinberg}}},\ }\href {\doibase 10.1086/151073} {\bibfield  {journal} {\bibinfo  {journal} {\apj}\ }\textbf {\bibinfo {volume} {168}},\ \bibinfo {pages} {175} (\bibinfo {year} {1971})}\BibitemShut {NoStop}%
\bibitem [{\citenamefont {{Udey}}\ and\ \citenamefont {{Israel}}(1982)}]{UdeyIsrael1982}%
  \BibitemOpen
  \bibfield  {author} {\bibinfo {author} {\bibfnamefont {N.}~\bibnamefont {{Udey}}}\ and\ \bibinfo {author} {\bibfnamefont {W.}~\bibnamefont {{Israel}}},\ }\href {\doibase 10.1093/mnras/199.4.1137} {\bibfield  {journal} {\bibinfo  {journal} {\mnras}\ }\textbf {\bibinfo {volume} {199}},\ \bibinfo {pages} {1137} (\bibinfo {year} {1982})}\BibitemShut {NoStop}%
\bibitem [{\citenamefont {{Thorne}}(1981)}]{Thorne1981}%
  \BibitemOpen
  \bibfield  {author} {\bibinfo {author} {\bibfnamefont {K.~S.}\ \bibnamefont {{Thorne}}},\ }\href {\doibase 10.1093/mnras/194.2.439} {\bibfield  {journal} {\bibinfo  {journal} {\mnras}\ }\textbf {\bibinfo {volume} {194}},\ \bibinfo {pages} {439} (\bibinfo {year} {1981})}\BibitemShut {NoStop}%
\bibitem [{\citenamefont {Anile}\ \emph {et~al.}(1992)\citenamefont {Anile}, \citenamefont {Pennisi},\ and\ \citenamefont {Sammartino}}]{AnileRadiazion1992}%
  \BibitemOpen
  \bibfield  {author} {\bibinfo {author} {\bibfnamefont {A.~M.}\ \bibnamefont {Anile}}, \bibinfo {author} {\bibfnamefont {S.}~\bibnamefont {Pennisi}}, \ and\ \bibinfo {author} {\bibfnamefont {M.}~\bibnamefont {Sammartino}},\ }\href {www.numdam.org/item/AIHPA_1992__56_1_49_0/} {\bibfield  {journal} {\bibinfo  {journal} {Annales de l'I.H.P. Physique th\'eorique}\ }\textbf {\bibinfo {volume} {56}},\ \bibinfo {pages} {49} (\bibinfo {year} {1992})}\BibitemShut {NoStop}%
\bibitem [{\citenamefont {{Farris}}\ \emph {et~al.}(2008)\citenamefont {{Farris}}, \citenamefont {{Li}}, \citenamefont {{Liu}},\ and\ \citenamefont {{Shapiro}}}]{Farris2008}%
  \BibitemOpen
  \bibfield  {author} {\bibinfo {author} {\bibfnamefont {B.~D.}\ \bibnamefont {{Farris}}}, \bibinfo {author} {\bibfnamefont {T.~K.}\ \bibnamefont {{Li}}}, \bibinfo {author} {\bibfnamefont {Y.~T.}\ \bibnamefont {{Liu}}}, \ and\ \bibinfo {author} {\bibfnamefont {S.~L.}\ \bibnamefont {{Shapiro}}},\ }\href {\doibase 10.1103/PhysRevD.78.024023} {\bibfield  {journal} {\bibinfo  {journal} {\prd}\ }\textbf {\bibinfo {volume} {78}},\ \bibinfo {eid} {024023} (\bibinfo {year} {2008})},\ \Eprint {http://arxiv.org/abs/0802.3210} {arXiv:0802.3210 [astro-ph]} \BibitemShut {NoStop}%
\bibitem [{\citenamefont {{Sadowski}}\ \emph {et~al.}(2013)\citenamefont {{Sadowski}}, \citenamefont {{Narayan}}, \citenamefont {{Tchekhovskoy}},\ and\ \citenamefont {{Zhu}}}]{Sadowski2013}%
  \BibitemOpen
  \bibfield  {author} {\bibinfo {author} {\bibfnamefont {A.}~\bibnamefont {{Sadowski}}}, \bibinfo {author} {\bibfnamefont {R.}~\bibnamefont {{Narayan}}}, \bibinfo {author} {\bibfnamefont {A.}~\bibnamefont {{Tchekhovskoy}}}, \ and\ \bibinfo {author} {\bibfnamefont {Y.}~\bibnamefont {{Zhu}}},\ }\href {\doibase 10.1093/mnras/sts632} {\bibfield  {journal} {\bibinfo  {journal} {\mnras}\ }\textbf {\bibinfo {volume} {429}},\ \bibinfo {pages} {3533} (\bibinfo {year} {2013})},\ \Eprint {http://arxiv.org/abs/1212.5050} {arXiv:1212.5050 [astro-ph.HE]} \BibitemShut {NoStop}%
\bibitem [{\citenamefont {de~Groot}\ \emph {et~al.}(1980)\citenamefont {de~Groot}, \citenamefont {van Leeuwen},\ and\ \citenamefont {van Weert}}]{Groot1980RelativisticKT}%
  \BibitemOpen
  \bibfield  {author} {\bibinfo {author} {\bibfnamefont {S.~R.}\ \bibnamefont {de~Groot}}, \bibinfo {author} {\bibfnamefont {W.~A.}\ \bibnamefont {van Leeuwen}}, \ and\ \bibinfo {author} {\bibfnamefont {C.~G.}\ \bibnamefont {van Weert}},\ }\href@noop {} {\emph {\bibinfo {title} {Relativistic kinetic theory: principles and applications}}}\ (\bibinfo {year} {1980})\BibitemShut {NoStop}%
\bibitem [{\citenamefont {{Misner}}\ \emph {et~al.}(1973)\citenamefont {{Misner}}, \citenamefont {{Thorne}},\ and\ \citenamefont {{Wheeler}}}]{MTW_book}%
  \BibitemOpen
  \bibfield  {author} {\bibinfo {author} {\bibfnamefont {C.~W.}\ \bibnamefont {{Misner}}}, \bibinfo {author} {\bibfnamefont {K.~S.}\ \bibnamefont {{Thorne}}}, \ and\ \bibinfo {author} {\bibfnamefont {J.~A.}\ \bibnamefont {{Wheeler}}},\ }\href@noop {} {\emph {\bibinfo {title} {{Gravitation}}}}\ (\bibinfo  {publisher} {W.H.~Freeman and Co.},\ \bibinfo {address} {San Francisco},\ \bibinfo {year} {1973})\BibitemShut {NoStop}%
\bibitem [{\citenamefont {{McLennan}}(1965)}]{McLennan1965}%
  \BibitemOpen
  \bibfield  {author} {\bibinfo {author} {\bibfnamefont {J.~A.}\ \bibnamefont {{McLennan}}},\ }\href {\doibase 10.1063/1.1761467} {\bibfield  {journal} {\bibinfo  {journal} {Physics of Fluids}\ }\textbf {\bibinfo {volume} {8}},\ \bibinfo {pages} {1580} (\bibinfo {year} {1965})}\BibitemShut {NoStop}%
\bibitem [{\citenamefont {Kurkela}\ and\ \citenamefont {Wiedemann}(2019)}]{Kurkela:2017xis}%
  \BibitemOpen
  \bibfield  {author} {\bibinfo {author} {\bibfnamefont {A.}~\bibnamefont {Kurkela}}\ and\ \bibinfo {author} {\bibfnamefont {U.~A.}\ \bibnamefont {Wiedemann}},\ }\href {\doibase 10.1140/epjc/s10052-019-7271-9} {\bibfield  {journal} {\bibinfo  {journal} {Eur. Phys. J. C}\ }\textbf {\bibinfo {volume} {79}},\ \bibinfo {pages} {776} (\bibinfo {year} {2019})},\ \Eprint {http://arxiv.org/abs/1712.04376} {arXiv:1712.04376 [hep-ph]} \BibitemShut {NoStop}%
\bibitem [{\citenamefont {{Grozdanov}}\ \emph {et~al.}(2019)\citenamefont {{Grozdanov}}, \citenamefont {{Lucas}},\ and\ \citenamefont {{Poovuttikul}}}]{Grozdanov2019}%
  \BibitemOpen
  \bibfield  {author} {\bibinfo {author} {\bibfnamefont {S.}~\bibnamefont {{Grozdanov}}}, \bibinfo {author} {\bibfnamefont {A.}~\bibnamefont {{Lucas}}}, \ and\ \bibinfo {author} {\bibfnamefont {N.}~\bibnamefont {{Poovuttikul}}},\ }\href {\doibase 10.1103/PhysRevD.99.086012} {\bibfield  {journal} {\bibinfo  {journal} {\prd}\ }\textbf {\bibinfo {volume} {99}},\ \bibinfo {eid} {086012} (\bibinfo {year} {2019})},\ \Eprint {http://arxiv.org/abs/1810.10016} {arXiv:1810.10016 [hep-th]} \BibitemShut {NoStop}%
\bibitem [{\citenamefont {Romatschke}(2016)}]{Romatschke:2015gic}%
  \BibitemOpen
  \bibfield  {author} {\bibinfo {author} {\bibfnamefont {P.}~\bibnamefont {Romatschke}},\ }\href {\doibase 10.1140/epjc/s10052-016-4169-7} {\bibfield  {journal} {\bibinfo  {journal} {Eur. Phys. J. C}\ }\textbf {\bibinfo {volume} {76}},\ \bibinfo {pages} {352} (\bibinfo {year} {2016})},\ \Eprint {http://arxiv.org/abs/1512.02641} {arXiv:1512.02641 [hep-th]} \BibitemShut {NoStop}%
\bibitem [{\citenamefont {Gavassino}\ \emph {et~al.}(2022)\citenamefont {Gavassino}, \citenamefont {Antonelli},\ and\ \citenamefont {Haskell}}]{GavassinoNonHydro2022}%
  \BibitemOpen
  \bibfield  {author} {\bibinfo {author} {\bibfnamefont {L.}~\bibnamefont {Gavassino}}, \bibinfo {author} {\bibfnamefont {M.}~\bibnamefont {Antonelli}}, \ and\ \bibinfo {author} {\bibfnamefont {B.}~\bibnamefont {Haskell}},\ }\href {\doibase 10.1103/PhysRevD.106.056010} {\bibfield  {journal} {\bibinfo  {journal} {Phys. Rev. D}\ }\textbf {\bibinfo {volume} {106}},\ \bibinfo {pages} {056010} (\bibinfo {year} {2022})}\BibitemShut {NoStop}%
\bibitem [{\citenamefont {Gavassino}(2024{\natexlab{a}})}]{GavassinoGapless2024rck}%
  \BibitemOpen
  \bibfield  {author} {\bibinfo {author} {\bibfnamefont {L.}~\bibnamefont {Gavassino}},\ }\href {\doibase 10.1103/PhysRevResearch.6.L042043} {\bibfield  {journal} {\bibinfo  {journal} {Phys. Rev. Res.}\ }\textbf {\bibinfo {volume} {6}},\ \bibinfo {pages} {L042043} (\bibinfo {year} {2024}{\natexlab{a}})},\ \Eprint {http://arxiv.org/abs/2404.12327} {arXiv:2404.12327 [nucl-th]} \BibitemShut {NoStop}%
\bibitem [{\citenamefont {Gavassino}(2024{\natexlab{b}})}]{GavassinoChapmanEnskog2024xwf}%
  \BibitemOpen
  \bibfield  {author} {\bibinfo {author} {\bibfnamefont {L.}~\bibnamefont {Gavassino}},\ }\href {\doibase 10.1103/PhysRevD.110.094012} {\bibfield  {journal} {\bibinfo  {journal} {Phys. Rev. D}\ }\textbf {\bibinfo {volume} {110}},\ \bibinfo {pages} {094012} (\bibinfo {year} {2024}{\natexlab{b}})},\ \Eprint {http://arxiv.org/abs/2408.14316} {arXiv:2408.14316 [nucl-th]} \BibitemShut {NoStop}%
\bibitem [{\citenamefont {{Spiegel}}(1957)}]{Spiegel1957}%
  \BibitemOpen
  \bibfield  {author} {\bibinfo {author} {\bibfnamefont {E.~A.}\ \bibnamefont {{Spiegel}}},\ }\href {\doibase 10.1086/146386} {\bibfield  {journal} {\bibinfo  {journal} {\apj}\ }\textbf {\bibinfo {volume} {126}},\ \bibinfo {pages} {202} (\bibinfo {year} {1957})}\BibitemShut {NoStop}%
\bibitem [{\citenamefont {Eckart}(1940)}]{Eckart40}%
  \BibitemOpen
  \bibfield  {author} {\bibinfo {author} {\bibfnamefont {C.}~\bibnamefont {Eckart}},\ }\href {\doibase 10.1103/PhysRev.58.919} {\bibfield  {journal} {\bibinfo  {journal} {Phys. Rev.}\ }\textbf {\bibinfo {volume} {58}},\ \bibinfo {pages} {919} (\bibinfo {year} {1940})}\BibitemShut {NoStop}%
\bibitem [{\citenamefont {Gavassino}\ \emph {et~al.}(2020{\natexlab{a}})\citenamefont {Gavassino}, \citenamefont {Antonelli},\ and\ \citenamefont {Haskell}}]{GavassinoLyapunov_2020}%
  \BibitemOpen
  \bibfield  {author} {\bibinfo {author} {\bibfnamefont {L.}~\bibnamefont {Gavassino}}, \bibinfo {author} {\bibfnamefont {M.}~\bibnamefont {Antonelli}}, \ and\ \bibinfo {author} {\bibfnamefont {B.}~\bibnamefont {Haskell}},\ }\href {\doibase 10.1103/physrevd.102.043018} {\bibfield  {journal} {\bibinfo  {journal} {Physical Review D}\ }\textbf {\bibinfo {volume} {102}} (\bibinfo {year} {2020}{\natexlab{a}}),\ 10.1103/physrevd.102.043018}\BibitemShut {NoStop}%
\bibitem [{\citenamefont {Israel}\ and\ \citenamefont {Stewart}(1979)}]{Israel_Stewart_1979}%
  \BibitemOpen
  \bibfield  {author} {\bibinfo {author} {\bibfnamefont {W.}~\bibnamefont {Israel}}\ and\ \bibinfo {author} {\bibfnamefont {J.}~\bibnamefont {Stewart}},\ }\href {\doibase https://doi.org/10.1016/0003-4916(79)90130-1} {\bibfield  {journal} {\bibinfo  {journal} {Annals of Physics}\ }\textbf {\bibinfo {volume} {118}},\ \bibinfo {pages} {341 } (\bibinfo {year} {1979})}\BibitemShut {NoStop}%
\bibitem [{\citenamefont {{Becattini}}(2016)}]{BecattiniBeta2016}%
  \BibitemOpen
  \bibfield  {author} {\bibinfo {author} {\bibfnamefont {F.}~\bibnamefont {{Becattini}}},\ }\href {\doibase 10.5506/APhysPolB.47.1819} {\bibfield  {journal} {\bibinfo  {journal} {Acta Physica Polonica B}\ }\textbf {\bibinfo {volume} {47}},\ \bibinfo {pages} {1819} (\bibinfo {year} {2016})},\ \Eprint {http://arxiv.org/abs/1606.06605} {arXiv:1606.06605 [gr-qc]} \BibitemShut {NoStop}%
\bibitem [{\citenamefont {Gavassino}(2020)}]{GavassinoTermometri}%
  \BibitemOpen
  \bibfield  {author} {\bibinfo {author} {\bibfnamefont {L.}~\bibnamefont {Gavassino}},\ }\href {\doibase 10.1007/s10701-020-00393-x} {\bibfield  {journal} {\bibinfo  {journal} {Found. Phys.}\ }\textbf {\bibinfo {volume} {50}},\ \bibinfo {pages} {1554} (\bibinfo {year} {2020})},\ \Eprint {http://arxiv.org/abs/2005.06396} {arXiv:2005.06396 [gr-qc]} \BibitemShut {NoStop}%
\bibitem [{\citenamefont {Gavassino}(2024{\natexlab{c}})}]{GavassinoShearRadiation2024jej}%
  \BibitemOpen
  \bibfield  {author} {\bibinfo {author} {\bibfnamefont {L.}~\bibnamefont {Gavassino}},\ }\href@noop {} {\  (\bibinfo {year} {2024}{\natexlab{c}})},\ \Eprint {http://arxiv.org/abs/2411.12929} {arXiv:2411.12929 [astro-ph.HE]} \BibitemShut {NoStop}%
\bibitem [{\citenamefont {{Heller}}\ \emph {et~al.}(2023)\citenamefont {{Heller}}, \citenamefont {{Serantes}}, \citenamefont {{Spali{\'n}ski}},\ and\ \citenamefont {{Withers}}}]{HellerBounds2023}%
  \BibitemOpen
  \bibfield  {author} {\bibinfo {author} {\bibfnamefont {M.~P.}\ \bibnamefont {{Heller}}}, \bibinfo {author} {\bibfnamefont {A.}~\bibnamefont {{Serantes}}}, \bibinfo {author} {\bibfnamefont {M.}~\bibnamefont {{Spali{\'n}ski}}}, \ and\ \bibinfo {author} {\bibfnamefont {B.}~\bibnamefont {{Withers}}},\ }\href {\doibase 10.1103/PhysRevLett.130.261601} {\bibfield  {journal} {\bibinfo  {journal} {\prl}\ }\textbf {\bibinfo {volume} {130}},\ \bibinfo {eid} {261601} (\bibinfo {year} {2023})},\ \Eprint {http://arxiv.org/abs/2212.07434} {arXiv:2212.07434 [hep-th]} \BibitemShut {NoStop}%
\bibitem [{\citenamefont {Hiscock}\ and\ \citenamefont {Lindblom}(1983)}]{Hishcock1983}%
  \BibitemOpen
  \bibfield  {author} {\bibinfo {author} {\bibfnamefont {W.~A.}\ \bibnamefont {Hiscock}}\ and\ \bibinfo {author} {\bibfnamefont {L.}~\bibnamefont {Lindblom}},\ }\href {\doibase https://doi.org/10.1016/0003-4916(83)90288-9} {\bibfield  {journal} {\bibinfo  {journal} {Annals of Physics}\ }\textbf {\bibinfo {volume} {151}},\ \bibinfo {pages} {466 } (\bibinfo {year} {1983})}\BibitemShut {NoStop}%
\bibitem [{\citenamefont {Landau}\ and\ \citenamefont {Lifshitz}(1987)}]{landau6}%
  \BibitemOpen
  \bibfield  {author} {\bibinfo {author} {\bibfnamefont {L.}~\bibnamefont {Landau}}\ and\ \bibinfo {author} {\bibfnamefont {E.}~\bibnamefont {Lifshitz}},\ }\href@noop {} {\emph {\bibinfo {title} {Fluid Mechanics}}},\ \bibinfo {number} {v. 6, Second Edition}\ (\bibinfo  {publisher} {Pergamon Press},\ \bibinfo {year} {1987})\BibitemShut {NoStop}%
\bibitem [{\citenamefont {Landau}\ and\ \citenamefont {Lifshitz}(1970)}]{landau7}%
  \BibitemOpen
  \bibfield  {author} {\bibinfo {author} {\bibfnamefont {L.}~\bibnamefont {Landau}}\ and\ \bibinfo {author} {\bibfnamefont {E.}~\bibnamefont {Lifshitz}},\ }\href@noop {} {\emph {\bibinfo {title} {Theory of elasticity}}},\ \bibinfo {number} {v. 7}\ (\bibinfo  {publisher} {Pergamon Press},\ \bibinfo {year} {1970})\BibitemShut {NoStop}%
\bibitem [{\citenamefont {{Rezzolla}}\ and\ \citenamefont {{Zanotti}}(2013)}]{rezzolla_book}%
  \BibitemOpen
  \bibfield  {author} {\bibinfo {author} {\bibfnamefont {L.}~\bibnamefont {{Rezzolla}}}\ and\ \bibinfo {author} {\bibfnamefont {O.}~\bibnamefont {{Zanotti}}},\ }\href@noop {} {\emph {\bibinfo {title} {Relativistic Hydrodynamics, by L.~Rezzolla and O.~Zanotti.~Oxford University Press, 2013.~ISBN-10: 0198528906; ISBN-13: 978-0198528906}}}\ (\bibinfo {year} {2013})\BibitemShut {NoStop}%
\bibitem [{\citenamefont {{Misner}}(1968)}]{Misner1968}%
  \BibitemOpen
  \bibfield  {author} {\bibinfo {author} {\bibfnamefont {C.~W.}\ \bibnamefont {{Misner}}},\ }\href {\doibase 10.1086/149448} {\bibfield  {journal} {\bibinfo  {journal} {\apj}\ }\textbf {\bibinfo {volume} {151}},\ \bibinfo {pages} {431} (\bibinfo {year} {1968})}\BibitemShut {NoStop}%
\bibitem [{\citenamefont {{Rebetzky}}\ \emph {et~al.}(1990)\citenamefont {{Rebetzky}}, \citenamefont {{Herold}}, \citenamefont {{Morfill}},\ and\ \citenamefont {{Ruder}}}]{Rebetzky1990}%
  \BibitemOpen
  \bibfield  {author} {\bibinfo {author} {\bibfnamefont {A.}~\bibnamefont {{Rebetzky}}}, \bibinfo {author} {\bibfnamefont {H.}~\bibnamefont {{Herold}}}, \bibinfo {author} {\bibfnamefont {G.}~\bibnamefont {{Morfill}}}, \ and\ \bibinfo {author} {\bibfnamefont {H.}~\bibnamefont {{Ruder}}},\ }\href {\doibase 10.1086/168431} {\bibfield  {journal} {\bibinfo  {journal} {\apj}\ }\textbf {\bibinfo {volume} {350}},\ \bibinfo {pages} {796} (\bibinfo {year} {1990})}\BibitemShut {NoStop}%
\bibitem [{\citenamefont {{Novikov}}\ and\ \citenamefont {{Thorne}}(1973)}]{NovikovThorne1973}%
  \BibitemOpen
  \bibfield  {author} {\bibinfo {author} {\bibfnamefont {I.~D.}\ \bibnamefont {{Novikov}}}\ and\ \bibinfo {author} {\bibfnamefont {K.~S.}\ \bibnamefont {{Thorne}}},\ }in\ \href@noop {} {\emph {\bibinfo {booktitle} {Black Holes (Les Astres Occlus)}}},\ \bibinfo {editor} {edited by\ \bibinfo {editor} {\bibfnamefont {C.}~\bibnamefont {{Dewitt}}}\ and\ \bibinfo {editor} {\bibfnamefont {B.~S.}\ \bibnamefont {{Dewitt}}}}\ (\bibinfo {year} {1973})\ pp.\ \bibinfo {pages} {343--450}\BibitemShut {NoStop}%
\bibitem [{\citenamefont {Hiscock}\ and\ \citenamefont {Lindblom}(1985)}]{Hiscock_Insatibility_first_order}%
  \BibitemOpen
  \bibfield  {author} {\bibinfo {author} {\bibfnamefont {W.}~\bibnamefont {Hiscock}}\ and\ \bibinfo {author} {\bibfnamefont {L.}~\bibnamefont {Lindblom}},\ }\href {\doibase 10.1103/PhysRevD.31.725} {\bibfield  {journal} {\bibinfo  {journal} {Physical review D: Particles and fields}\ }\textbf {\bibinfo {volume} {31}},\ \bibinfo {pages} {725} (\bibinfo {year} {1985})}\BibitemShut {NoStop}%
\bibitem [{\citenamefont {Gavassino}(2022)}]{GavassinoSuperluminal2021}%
  \BibitemOpen
  \bibfield  {author} {\bibinfo {author} {\bibfnamefont {L.}~\bibnamefont {Gavassino}},\ }\href {\doibase 10.1103/PhysRevX.12.041001} {\bibfield  {journal} {\bibinfo  {journal} {Phys. Rev. X}\ }\textbf {\bibinfo {volume} {12}},\ \bibinfo {pages} {041001} (\bibinfo {year} {2022})}\BibitemShut {NoStop}%
\bibitem [{\citenamefont {Gavassino}(2023)}]{GavassinoBounds2023}%
  \BibitemOpen
  \bibfield  {author} {\bibinfo {author} {\bibfnamefont {L.}~\bibnamefont {Gavassino}},\ }\href {\doibase 10.1016/j.physletb.2023.137854} {\bibfield  {journal} {\bibinfo  {journal} {Phys. Lett. B}\ }\textbf {\bibinfo {volume} {840}},\ \bibinfo {pages} {137854} (\bibinfo {year} {2023})},\ \Eprint {http://arxiv.org/abs/2301.06651} {arXiv:2301.06651 [hep-th]} \BibitemShut {NoStop}%
\bibitem [{\citenamefont {Gavassino}\ \emph {et~al.}(2024)\citenamefont {Gavassino}, \citenamefont {Disconzi},\ and\ \citenamefont {Noronha}}]{GavassinoDispersion2024}%
  \BibitemOpen
  \bibfield  {author} {\bibinfo {author} {\bibfnamefont {L.}~\bibnamefont {Gavassino}}, \bibinfo {author} {\bibfnamefont {M.}~\bibnamefont {Disconzi}}, \ and\ \bibinfo {author} {\bibfnamefont {J.}~\bibnamefont {Noronha}},\ }\href {\doibase 10.1103/PhysRevLett.132.162301} {\bibfield  {journal} {\bibinfo  {journal} {Phys. Rev. Lett.}\ }\textbf {\bibinfo {volume} {132}},\ \bibinfo {pages} {162301} (\bibinfo {year} {2024})}\BibitemShut {NoStop}%
\bibitem [{\citenamefont {H\"{o}rmander}(1989)}]{HormanderOperatorsBook}%
  \BibitemOpen
  \bibfield  {author} {\bibinfo {author} {\bibfnamefont {L.}~\bibnamefont {H\"{o}rmander}},\ }\href@noop {} {\emph {\bibinfo {title} {The analysis of Linear Partial Differential Operators I (second edition)}}}\ (\bibinfo  {publisher} {Springer-Verlag},\ \bibinfo {address} {Berlin},\ \bibinfo {year} {1989})\BibitemShut {NoStop}%
\bibitem [{\citenamefont {Evans}(1997)}]{EvansPDEsBook}%
  \BibitemOpen
  \bibfield  {author} {\bibinfo {author} {\bibfnamefont {L.~C.}\ \bibnamefont {Evans}},\ }\href@noop {} {\emph {\bibinfo {title} {Partial Differential Equations}}}\ (\bibinfo  {publisher} {American Mathematical Society},\ \bibinfo {address} {Berkeley},\ \bibinfo {year} {1997})\BibitemShut {NoStop}%
\bibitem [{\citenamefont {Cattaneo}(1958)}]{cattaneo1958}%
  \BibitemOpen
  \bibfield  {author} {\bibinfo {author} {\bibfnamefont {C.}~\bibnamefont {Cattaneo}},\ }\href {https://books.google.pl/books?id=mHGeQwAACAAJ} {\emph {\bibinfo {title} {Sur une forme de l'{\'e}quation de la chaleur {\'e}liminant le paradoxe d'une propagation instantan{\'e}e}}},\ Comptes rendus hebdomadaires des s{\'e}ances de l'Acad{\'e}mie des sciences\ (\bibinfo  {publisher} {Gauthier-Villars},\ \bibinfo {year} {1958})\BibitemShut {NoStop}%
\bibitem [{\citenamefont {{Minerbo}}(1978)}]{Minerbo1978}%
  \BibitemOpen
  \bibfield  {author} {\bibinfo {author} {\bibfnamefont {G.~N.}\ \bibnamefont {{Minerbo}}},\ }\href {\doibase 10.1016/0022-4073(78)90024-9} {\bibfield  {journal} {\bibinfo  {journal} {\jqsrt}\ }\textbf {\bibinfo {volume} {20}},\ \bibinfo {pages} {541} (\bibinfo {year} {1978})}\BibitemShut {NoStop}%
\bibitem [{\citenamefont {{Levermore}}(1984)}]{Levermore1984}%
  \BibitemOpen
  \bibfield  {author} {\bibinfo {author} {\bibfnamefont {C.~D.}\ \bibnamefont {{Levermore}}},\ }\href {\doibase 10.1016/0022-4073(84)90112-2} {\bibfield  {journal} {\bibinfo  {journal} {\jqsrt}\ }\textbf {\bibinfo {volume} {31}},\ \bibinfo {pages} {149} (\bibinfo {year} {1984})}\BibitemShut {NoStop}%
\bibitem [{\citenamefont {Fragile}\ \emph {et~al.}(2014)\citenamefont {Fragile}, \citenamefont {Olejar},\ and\ \citenamefont {Anninos}}]{Fragile:2014bfa}%
  \BibitemOpen
  \bibfield  {author} {\bibinfo {author} {\bibfnamefont {P.~C.}\ \bibnamefont {Fragile}}, \bibinfo {author} {\bibfnamefont {A.}~\bibnamefont {Olejar}}, \ and\ \bibinfo {author} {\bibfnamefont {P.}~\bibnamefont {Anninos}},\ }\href {\doibase 10.1088/0004-637X/796/1/22} {\bibfield  {journal} {\bibinfo  {journal} {Astrophys. J.}\ }\textbf {\bibinfo {volume} {796}},\ \bibinfo {pages} {22} (\bibinfo {year} {2014})},\ \Eprint {http://arxiv.org/abs/1408.4460} {arXiv:1408.4460 [astro-ph.IM]} \BibitemShut {NoStop}%
\bibitem [{\citenamefont {Fragile}\ \emph {et~al.}(2018)\citenamefont {Fragile}, \citenamefont {Etheridge}, \citenamefont {Anninos}, \citenamefont {Mishra},\ and\ \citenamefont {Kluzniak}}]{Fragile:2018xee}%
  \BibitemOpen
  \bibfield  {author} {\bibinfo {author} {\bibfnamefont {P.~C.}\ \bibnamefont {Fragile}}, \bibinfo {author} {\bibfnamefont {S.~M.}\ \bibnamefont {Etheridge}}, \bibinfo {author} {\bibfnamefont {P.}~\bibnamefont {Anninos}}, \bibinfo {author} {\bibfnamefont {B.}~\bibnamefont {Mishra}}, \ and\ \bibinfo {author} {\bibfnamefont {W.}~\bibnamefont {Kluzniak}},\ }\href {\doibase 10.3847/1538-4357/aab788} {\bibfield  {journal} {\bibinfo  {journal} {Astrophys. J.}\ }\textbf {\bibinfo {volume} {857}},\ \bibinfo {pages} {1} (\bibinfo {year} {2018})},\ \Eprint {http://arxiv.org/abs/1803.06423} {arXiv:1803.06423 [astro-ph.HE]} \BibitemShut {NoStop}%
\bibitem [{\citenamefont {Murchikova}\ \emph {et~al.}(2017)\citenamefont {Murchikova}, \citenamefont {Abdikamalov},\ and\ \citenamefont {Urbatsch}}]{Murchikova:2017zsy}%
  \BibitemOpen
  \bibfield  {author} {\bibinfo {author} {\bibfnamefont {L.~M.}\ \bibnamefont {Murchikova}}, \bibinfo {author} {\bibfnamefont {E.}~\bibnamefont {Abdikamalov}}, \ and\ \bibinfo {author} {\bibfnamefont {T.}~\bibnamefont {Urbatsch}},\ }\href {\doibase 10.1093/mnras/stx986} {\bibfield  {journal} {\bibinfo  {journal} {Mon. Not. Roy. Astron. Soc.}\ }\textbf {\bibinfo {volume} {469}},\ \bibinfo {pages} {1725} (\bibinfo {year} {2017})},\ \Eprint {http://arxiv.org/abs/1701.07027} {arXiv:1701.07027 [astro-ph.HE]} \BibitemShut {NoStop}%
\bibitem [{\citenamefont {{Anninos}}\ and\ \citenamefont {{Fragile}}(2020)}]{Anninos2020}%
  \BibitemOpen
  \bibfield  {author} {\bibinfo {author} {\bibfnamefont {P.}~\bibnamefont {{Anninos}}}\ and\ \bibinfo {author} {\bibfnamefont {P.~C.}\ \bibnamefont {{Fragile}}},\ }\href {\doibase 10.3847/1538-4357/abab9c} {\bibfield  {journal} {\bibinfo  {journal} {\apj}\ }\textbf {\bibinfo {volume} {900}},\ \bibinfo {eid} {71} (\bibinfo {year} {2020})},\ \Eprint {http://arxiv.org/abs/2007.12195} {arXiv:2007.12195 [astro-ph.IM]} \BibitemShut {NoStop}%
\bibitem [{\citenamefont {Gavassino}\ \emph {et~al.}(2020{\natexlab{b}})\citenamefont {Gavassino}, \citenamefont {Antonelli},\ and\ \citenamefont {Haskell}}]{GavassinoRadiazione}%
  \BibitemOpen
  \bibfield  {author} {\bibinfo {author} {\bibfnamefont {L.}~\bibnamefont {Gavassino}}, \bibinfo {author} {\bibfnamefont {M.}~\bibnamefont {Antonelli}}, \ and\ \bibinfo {author} {\bibfnamefont {B.}~\bibnamefont {Haskell}},\ }\href {\doibase 10.3390/sym12091543} {\bibfield  {journal} {\bibinfo  {journal} {Symmetry}\ }\textbf {\bibinfo {volume} {12}},\ \bibinfo {pages} {1543} (\bibinfo {year} {2020}{\natexlab{b}})}\BibitemShut {NoStop}%
\bibitem [{\citenamefont {{Anderson}}\ and\ \citenamefont {{Spiegel}}(1972)}]{AndersonSpiegel1972}%
  \BibitemOpen
  \bibfield  {author} {\bibinfo {author} {\bibfnamefont {J.~L.}\ \bibnamefont {{Anderson}}}\ and\ \bibinfo {author} {\bibfnamefont {E.~A.}\ \bibnamefont {{Spiegel}}},\ }\href {\doibase 10.1086/151265} {\bibfield  {journal} {\bibinfo  {journal} {\apj}\ }\textbf {\bibinfo {volume} {171}},\ \bibinfo {pages} {127} (\bibinfo {year} {1972})}\BibitemShut {NoStop}%
\bibitem [{\citenamefont {Gavassino}\ and\ \citenamefont {Antonelli}(2021)}]{GavassinoFronntiers2021}%
  \BibitemOpen
  \bibfield  {author} {\bibinfo {author} {\bibfnamefont {L.}~\bibnamefont {Gavassino}}\ and\ \bibinfo {author} {\bibfnamefont {M.}~\bibnamefont {Antonelli}},\ }\href {\doibase 10.3389/fspas.2021.686344} {\bibfield  {journal} {\bibinfo  {journal} {Front. Astron. Space Sci.}\ }\textbf {\bibinfo {volume} {8}},\ \bibinfo {pages} {686344} (\bibinfo {year} {2021})},\ \Eprint {http://arxiv.org/abs/2105.15184} {arXiv:2105.15184 [gr-qc]} \BibitemShut {NoStop}%
\bibitem [{\citenamefont {Moore}(2018)}]{MooreCuts2018mma}%
  \BibitemOpen
  \bibfield  {author} {\bibinfo {author} {\bibfnamefont {G.~D.}\ \bibnamefont {Moore}},\ }\href {\doibase 10.1007/JHEP05(2018)084} {\bibfield  {journal} {\bibinfo  {journal} {JHEP}\ }\textbf {\bibinfo {volume} {05}},\ \bibinfo {pages} {084} (\bibinfo {year} {2018})},\ \Eprint {http://arxiv.org/abs/1803.00736} {arXiv:1803.00736 [hep-ph]} \BibitemShut {NoStop}%
\bibitem [{\citenamefont {{Romatschke}}\ and\ \citenamefont {{Romatschke}}(2017)}]{Romatschke2017}%
  \BibitemOpen
  \bibfield  {author} {\bibinfo {author} {\bibfnamefont {P.}~\bibnamefont {{Romatschke}}}\ and\ \bibinfo {author} {\bibfnamefont {U.}~\bibnamefont {{Romatschke}}},\ }\href {\doibase 10.48550/arXiv.1712.05815} {\bibfield  {journal} {\bibinfo  {journal} {arXiv e-prints}\ ,\ \bibinfo {eid} {arXiv:1712.05815}} (\bibinfo {year} {2017})},\ \Eprint {http://arxiv.org/abs/1712.05815} {arXiv:1712.05815 [nucl-th]} \BibitemShut {NoStop}%
\bibitem [{\citenamefont {{Florkowski}}\ \emph {et~al.}(2018)\citenamefont {{Florkowski}}, \citenamefont {{Heller}},\ and\ \citenamefont {{Spali{\'n}ski}}}]{FlorkowskiReview2018}%
  \BibitemOpen
  \bibfield  {author} {\bibinfo {author} {\bibfnamefont {W.}~\bibnamefont {{Florkowski}}}, \bibinfo {author} {\bibfnamefont {M.~P.}\ \bibnamefont {{Heller}}}, \ and\ \bibinfo {author} {\bibfnamefont {M.}~\bibnamefont {{Spali{\'n}ski}}},\ }\href {\doibase 10.1088/1361-6633/aaa091} {\bibfield  {journal} {\bibinfo  {journal} {Reports on Progress in Physics}\ }\textbf {\bibinfo {volume} {81}},\ \bibinfo {eid} {046001} (\bibinfo {year} {2018})},\ \Eprint {http://arxiv.org/abs/1707.02282} {arXiv:1707.02282 [hep-ph]} \BibitemShut {NoStop}%
\bibitem [{\citenamefont {Rocha}\ \emph {et~al.}(2023)\citenamefont {Rocha}, \citenamefont {Wagner}, \citenamefont {Denicol}, \citenamefont {Noronha},\ and\ \citenamefont {Rischke}}]{RochaReview2023ilf}%
  \BibitemOpen
  \bibfield  {author} {\bibinfo {author} {\bibfnamefont {G.~S.}\ \bibnamefont {Rocha}}, \bibinfo {author} {\bibfnamefont {D.}~\bibnamefont {Wagner}}, \bibinfo {author} {\bibfnamefont {G.~S.}\ \bibnamefont {Denicol}}, \bibinfo {author} {\bibfnamefont {J.}~\bibnamefont {Noronha}}, \ and\ \bibinfo {author} {\bibfnamefont {D.~H.}\ \bibnamefont {Rischke}},\ }\href@noop {} {\  (\bibinfo {year} {2023})},\ \Eprint {http://arxiv.org/abs/2311.15063} {arXiv:2311.15063 [nucl-th]} \BibitemShut {NoStop}%
\bibitem [{\citenamefont {{Dudy{\'n}ski}}(1989)}]{Dudynski1989}%
  \BibitemOpen
  \bibfield  {author} {\bibinfo {author} {\bibfnamefont {M.}~\bibnamefont {{Dudy{\'n}ski}}},\ }\href {\doibase 10.1007/BF01023641} {\bibfield  {journal} {\bibinfo  {journal} {Journal of Statistical Physics}\ }\textbf {\bibinfo {volume} {57}},\ \bibinfo {pages} {199} (\bibinfo {year} {1989})}\BibitemShut {NoStop}%
\bibitem [{\citenamefont {Struchtrup}\ and\ \citenamefont {Taheri}(2011)}]{Struchtrup2011}%
  \BibitemOpen
  \bibfield  {author} {\bibinfo {author} {\bibfnamefont {H.}~\bibnamefont {Struchtrup}}\ and\ \bibinfo {author} {\bibfnamefont {P.}~\bibnamefont {Taheri}},\ }\href {\doibase 10.1093/imamat/hxr004} {\bibfield  {journal} {\bibinfo  {journal} {IMA Journal of Applied Mathematics}\ }\textbf {\bibinfo {volume} {76}},\ \bibinfo {pages} {672} (\bibinfo {year} {2011})},\ \Eprint {http://arxiv.org/abs/https://academic.oup.com/imamat/article-pdf/76/5/672/1835580/hxr004.pdf} {https://academic.oup.com/imamat/article-pdf/76/5/672/1835580/hxr004.pdf} \BibitemShut {NoStop}%
\bibitem [{\citenamefont {Heller}\ \emph {et~al.}(2024)\citenamefont {Heller}, \citenamefont {Serantes}, \citenamefont {Spali\'nski},\ and\ \citenamefont {Withers}}]{HellerHydroHedron2023jtd}%
  \BibitemOpen
  \bibfield  {author} {\bibinfo {author} {\bibfnamefont {M.~P.}\ \bibnamefont {Heller}}, \bibinfo {author} {\bibfnamefont {A.}~\bibnamefont {Serantes}}, \bibinfo {author} {\bibfnamefont {M.}~\bibnamefont {Spali\'nski}}, \ and\ \bibinfo {author} {\bibfnamefont {B.}~\bibnamefont {Withers}},\ }\href {https://doi.org/10.1038/s41567-024-02635-5} {\bibfield  {journal} {\bibinfo  {journal} {Nature Physics}\ } (\bibinfo {year} {2024})}\BibitemShut {NoStop}%
\bibitem [{\citenamefont {{Peliti}}(2011)}]{peliti_book}%
  \BibitemOpen
  \bibfield  {author} {\bibinfo {author} {\bibfnamefont {L.}~\bibnamefont {{Peliti}}},\ }\href@noop {} {\emph {\bibinfo {title} {{Statistical Mechanics in a Nutshell}}}},\ In a nutshell\ (\bibinfo  {publisher} {Princeton University Press},\ \bibinfo {year} {2011})\BibitemShut {NoStop}%
\bibitem [{\citenamefont {Czajka}\ and\ \citenamefont {Jeon}(2017)}]{Czajka:2017bod}%
  \BibitemOpen
  \bibfield  {author} {\bibinfo {author} {\bibfnamefont {A.}~\bibnamefont {Czajka}}\ and\ \bibinfo {author} {\bibfnamefont {S.}~\bibnamefont {Jeon}},\ }\href {\doibase 10.1103/PhysRevC.95.064906} {\bibfield  {journal} {\bibinfo  {journal} {Phys. Rev. C}\ }\textbf {\bibinfo {volume} {95}},\ \bibinfo {pages} {064906} (\bibinfo {year} {2017})},\ \Eprint {http://arxiv.org/abs/1701.07580} {arXiv:1701.07580 [nucl-th]} \BibitemShut {NoStop}%
\bibitem [{\citenamefont {{Shavaliev}}(1993)}]{SuperBurnett1993}%
  \BibitemOpen
  \bibfield  {author} {\bibinfo {author} {\bibfnamefont {M.~S.}\ \bibnamefont {{Shavaliev}}},\ }\href@noop {} {\bibfield  {journal} {\bibinfo  {journal} {Prikladnaia Matematika i Mekhanika}\ }\textbf {\bibinfo {volume} {57}},\ \bibinfo {pages} {168} (\bibinfo {year} {1993})}\BibitemShut {NoStop}%
\end{thebibliography}%

\label{lastpage}

\end{document}